\documentclass[prl, amsmath, amssymb, reprint,
               superscriptaddress, showkeys]{revtex4-1}
\pdfoutput=1
\usepackage{amssymb,amsmath}

\usepackage{graphicx}

\usepackage{braket}

\usepackage[pdftex,
            pdfauthor={Josef Rammensee, Juan Diego Urbina, Klaus Richter},
            pdftitle={Many-Body Quantum Interference and the Saturation of
              Out-of-Time-Order Correlators}]{hyperref}

\renewcommand{\vec}[1]{\mathbf{#1}}  

\newcommand{\rmi}{\textrm{i}} 
\newcommand{\rmd}{\textrm{d}} 
\newcommand{\rme}{\textrm{e}} 
\newcommand{\rms}{\textrm{s}} 
\newcommand{\rmu}{\textrm{u}} 
\newcommand{\rmcl}{\textrm{cl}} 

\newcommand{\init}[1]{{#1}^{(\textrm{i})}}
\newcommand{\fin}[1]{{#1}^{(\textrm{f})}}

\newcommand{\vecfin}[1]{\fin{\vec{#1}}}
\newcommand{\vecinit}[1]{\init{\vec{#1}}}

\newcommand{\cop}[1]{\hat{b}_{#1}^\dagger}
\newcommand{\anop}[1]{\hat{b}_{#1}}

\newcommand{\tauE}{{\tau_{\mathrm{E}}}}    
\newcommand{\heff}{\hbar_{\mathrm{eff}}} 

\newcommand{\ts}{t_\rms}           
\newcommand{\tu}{t_\rmu}           
\newcommand{\tenc}{t_{\mathrm{enc}}} 

\newcommand{\diagram}[2]{\raisebox{-0.5\height}{\includegraphics[#1]{#2}}}

\def\?#1{}

\newcommand{\veces}[3][\?]{{\vec{e}_{#1,\rms}^{(#2)}\!(#3)}}
\newcommand{\veceu}[3][\?]{{\vec{e}_{#1,\rmu}^{(#2)}\!(#3)}}

\DeclareMathOperator{\Si}{Si} 

\begin{document}
%
%
\title{Many-Body Quantum Interference and the Saturation\\ of
  Out-of-Time-Order Correlators}
\newcommand{\RegensburgUniversity}{Institut f\"ur Theoretische Physik,
  Universit\"at Regensburg, D-93040 Regensburg, Germany}
\author{Josef Rammensee}
\affiliation{\RegensburgUniversity}
\author{Juan Diego Urbina}
\affiliation{\RegensburgUniversity}
\author{Klaus Richter}
\email{klaus.richter@physik.uni-regensburg.de}
\affiliation{\RegensburgUniversity}

\begin{abstract}
  Out-of-time-order correlators (OTOCs) have been proposed as
  sensitive probes for chaos in interacting quantum systems.
  They exhibit a characteristic classical exponential growth, but
  saturate beyond the so-called scrambling or Ehrenfest time $\tauE$
  in the quantum correlated regime.
  Here we present a path-integral approach for the entire time
  evolution of OTOCs for bosonic $N$-particle systems.
  We first show how the growth of OTOCs up to
  $\tauE\!=\!(1/\lambda)\log N$ is related to the Lyapunov exponent
  $\lambda$ of the corresponding chaotic mean-field dynamics in the
  semiclassical large-$N$ limit.
  Beyond $\tauE$, where simple mean-field approaches break down, we
  identify the underlying quantum mechanism responsible for the
  saturation.
  To this end we express OTOCs by coherent sums over contributions
  from different mean-field solutions and compute the dominant
  many-body interference term amongst them.
  Our method further applies to the complementary semiclassical limit
  $\hbar\! \rightarrow\! 0$ for fixed $N$, including quantum-chaotic
  single- and few-particle systems.
\end{abstract}

\maketitle

The study of signatures of unstable classical dynamics in the spectral
and dynamical properties of corresponding quantum systems, known as
quantum chaos~\cite{Gut1991Book}, has recently received particular
attention after the proposal of Kitaev \footnote{A. Kitaev,
  \emph{Hidden Correlations in the Hawking Radiation and Thermal
    Noise}, talk at Breakthrough Physics Prize Symposium, Nov. 10,
  2014, \url{https://www.youtube.com/watch?v=OQ9qN8j7EZI} } and
related works~\cite{SekSusJHEP2008,SheStaJHEP2014,MalSheSta2015} that
address the mechanisms for spreading or ``scrambling'' quantum
information across the many degrees of freedom of interacting
many-body (MB) systems.
With regard to such a MB quantum-to-classical correspondence,
out-of-time-order correlators (OTOCs)
\cite{LarOvcJETP1969,MalSheSta2015}, such as
\begin{equation}
  C(t)=\Braket{
  \left[ \hat{V}(t), \hat{W}(0)  \right]^\dagger
  \left[ \hat{V}(t),\hat{W}(0)  \right]
  }\,,
  \label{eq:OTOC_definition}
\end{equation}
are measures of choice (with several experimental protocols already
available~\cite{ZhuHafGroPRA2016, SwiBenSchleHazPRA2016,
  CamGooPRE2016, LiFanWanZeYenYhaPenDuPRX2016,
  GarBohSafWalBolRey2016}):
The squared commutator of a suitable (local) operator $\hat{V}(t)$
with another (local) perturbation $\hat{W}(0)$ probes the temporal
growth of $\hat{V}$, including its growing complexity.
Hence, due to their unusual time ordering, OTOCs represent MB quantum
analogues of classical measures for instability of chaotic MB
dynamics.
Indeed, invoking a heuristic classical-to-quantum correspondence for
small $\hbar$ and replacing the commutator in
Eq.~(\ref{eq:OTOC_definition}) for short times by Poisson brackets one
obtains, {\em e.g.}, for $\hat{W}\!=\!\hat{p}_i$,
$\hat{V}\!=\!\hat{q}_j$
\cite{LarOvcJETP1969,MalSheSta2015,SwiBenSchleHazPRA2016},
\begin{equation}
  |\rmi \hbar|^2 \Braket{\left\{\init{p}_i,\fin{q}_j(t) \right\}^{\!2}}
  \! = \! \hbar^2 \! \Braket{
  \left(\frac{\partial \fin{q}_j}{\partial \init{q}_i}(t)\!
  \right)^{\!2}} \propto
   \hbar^2 \rme^{2\lambda t} \, .
  \label{eq:OTOC_Moyal}
\end{equation}
Here the averages $\langle \cdots \rangle$ are taken over the initial
phase-space points $(\vec{q},\vec{p})$ weighted by the corresponding
quasidistribution.
The exponential growth on the rhs follows from the relation
$|\partial \fin{q}_j / \partial \init{q}_i| \!\propto\!  \rme^{\lambda
  t}$ for chaotic systems with average single-particle (SP) Lyapunov
exponent $\lambda$, see also Ref.~\cite{Kur2018JSP} for another
semiclassical derivation.
Intriguingly, in view of Eq.~(\ref{eq:OTOC_Moyal}), the genuinely
quantum-mechanical OTOC $C(t)$ provides a direct measure of classical
chaos in the corresponding quantum system, similar to the Loschmidt
echo \cite{JalPasPRL2001}.
This close correspondence has been unambiguously observed in numerical
studies for SP systems~\cite{RoGaGaPRL2017}.
For MB problems analytical works have focused on Sachdev-Ye-Kitaev models
\cite{BagAltKamNPB2017,ScaAltarxiv2017} or used random matrix theory
(where $\lambda\!\to\!\infty$) \cite{TorGarSan2018PRB,
  CotHunKLiuZosJHEP2017, DelMolSonPRD2017}, while the numerical
identification of a MB Lyapunov exponent from
Eq.~(\ref{eq:OTOC_definition}) remains a challenge
\cite{BohMenEndKnaNJP2016,SheZhaFanZhaPRB2016,HasMurYosJHEP2017}.

Moreover, Eq.~(\ref{eq:OTOC_Moyal}) predicts unbounded classical
growth while $C(t)$ is eventually bounded due to quantum mechanical
unitarity.
Indeed, $C(t)$ is numerically found \cite{RoGaGaPRL2017,
  BohMenEndKnaNJP2016} to saturate beyond a characteristic time scale,
known as Ehrenfest time $\tauE$~\cite{EhrZfP1927,BerZasPA1978} and
dubbed scrambling time~\cite{MalSheSta2015,DvaFlaGomPriPRD2013} in the
MB context.
$\tauE$ separates initial quantum evolution following essentially
classical motion from dynamics dominated by interference effects.
Accordingly, quantum interference has been assumed to cause saturation
of OTOCs in some way \cite{SekSusJHEP2008,HasMurYosJHEP2017,
  BagAltKamNPB2017, RoGaGaPRL2017}, but to date the precise underlying
dynamical mechanism has yet been unknown for chaotic SP and MB
systems.

This classical-to-quantum crossover happens at
$\tauE=(1/\lambda) \log (1/\heff)$ where
``$\heff \! \rightarrow \! 0$'' can denote complementary semiclassical
limits:
For fixed $N$, $\heff\! \sim\! \hbar$ and $\lambda$ is the
characteristic Lyapunov exponent of the limiting classical particle
dynamics [see Eq.~(\ref{eq:OTOC_Moyal}] for $N\!=\!1$).
For MB systems with a complementary classical, large-$N$ mean-field
limit, $\heff \!\simeq\! 1/N \! $ and $\lambda$ characterizes the
instability of the corresponding nonlinear mean-field solutions.

\begin{figure}[t]
  \includegraphics[width=\linewidth]{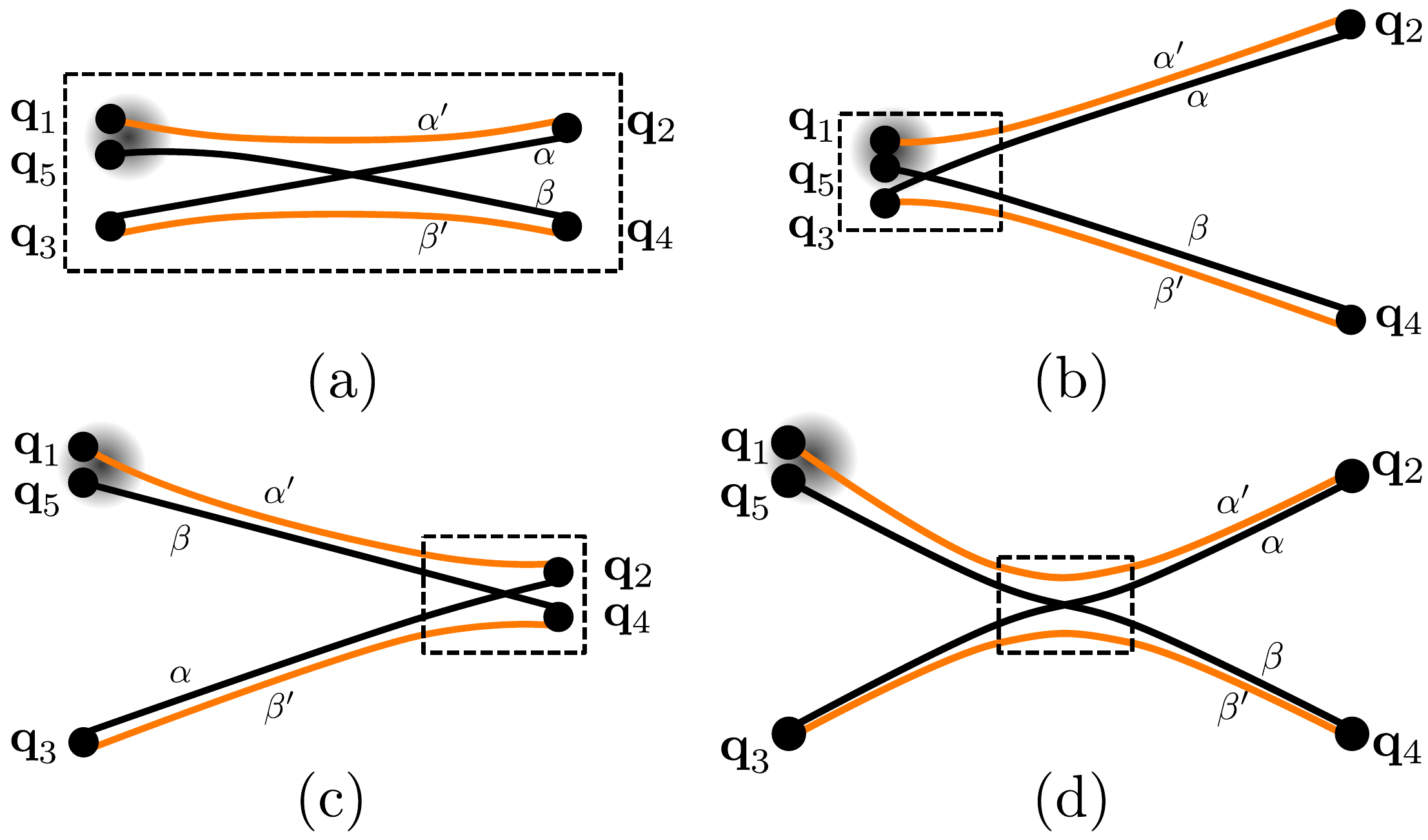}
     \caption{ Trajectory configurations representing interfering
       mean-field solutions that dominantly contribute to the OTOC
       $C(t)$, Eq.~(\ref{eq:OTOC_sc_integral_representation}).  The
       trajectory quadruples reside (a) inside an encounter (marked by
       dashed box), form a ``two-leg'' diagram with an encounter (b) at
       the beginning or (c) at the end, or (d) build a
       ``four-leg'' diagram with the encounter in between.}
    \label{fig:OTOC_diagram_classes}
\end{figure}

The notable interference-based saturation of OTOCs beyond $\tauE$ is
not captured by a Moyal expansion \cite{CotHunKLiuZosJHEP2017,
  ScaAltarxiv2017} of commutators [such as
Eq.~(\ref{eq:OTOC_definition})] in powers of $\heff$ as implicit in
Eq.~(\ref{eq:OTOC_Moyal}).
However, as originally developed for SP~\cite{TomHelPRL1991,
  AleLarPRB1996, AgaAleLarPRL2000, SieRicPS2001,
  MulHeuBraHaaPRE2005, BroRahPRB2006,
  JacWhiPRB2006, WalGutGouRicPRL2008} and recently extended to MB
systems \cite{EngDujArgSchlaRicUrb13PRL, UrbKuiHumMatRicPRL2016,
  EngUrbRicPRE2015, DubMulNJP2016, AkiWalGutBraGuhPRL2017,
  TomSchlaUllUrbRic2018PRA}, there exist semiclassical techniques
that adequately describe post-Ehrenfest quantum phenomena.
By extending these approaches to MB commutator norms, here we develop
a unifying semiclassical theory for OTOCs which bridges classical
mean-field and quantum MB concepts for bosonic large-$N$ systems. The
complementary limit ``$\hbar\! \rightarrow\! 0$'' for fixed $N$ will
be discussed at the end.
We express OTOCs through semiclassical propagators in Fock space
\cite{EngDujArgSchlaRicUrb13PRL} leading to sums over amplitudes from
unstable classical paths, {\em i.e.}, mean-field solutions.
By considering subtle classical correlations amongst them we identify
and compute the dominant contributions involving correlated MB
dynamics swapping forth and back between mean-field paths (see
Fig.~\ref{fig:OTOC_diagram_classes}).
They proof responsible for the initial exponential growth and the
saturation of OTOCs.

Specifically, we consider Bose-Hubbard systems with $n$ sites
describing $N$ interacting bosons with Hamiltonian
\begin{equation}
  \hat{H}
  =
  \sum_{ij=1}^n h_{ij} \cop{i} \anop{j}
  +\frac{1}{N} \sum_{ijkl=1}^n V_{ijkl} \cop{i} \cop{j} \anop{k} \anop{l}
  \, ,
  \label{eq:H}
\end{equation}
where $\cop{i}$ ($\anop{i}$) are creation (annihilation) operators at
sites $i\!=\!1,\ldots,n$.
The parameters $h_{ij}$ define on-site energies and hopping terms, and
$V_{ijkl}$ denote interactions.

We evaluate the OTOC Eq.~(\ref{eq:OTOC_definition}) for position and
momentum quadrature operators~\cite{Agarwal2013}
$ \hat{q}_i\! =\! ( \anop{i} \!+\! \cop{i}) / \sqrt{2N} \,,\,
\hat{p}_i \!=\!  ( \anop{i} \!-\! \cop{i} ) / ( \sqrt{2N} \rmi )$,
related to occupation operators $\hat{n}_i$ through
$ ( \hat{q}_i^2 \!+\! \hat{p}^2_i ) / 2 \!=\! \heff ( \hat{n}_i \!+\!
1 / 2 ) $.
Using the MB time evolution operator
$\hat{U}(t) \!=\! \exp(-\rmi \hat{H} t / \hbar)$
Eq.~(\ref{eq:OTOC_definition}) reads
\begin{equation}
  C(t)\!=\!\Braket{\Psi| \!
  \left[ \hat{p}_i,\hat{U}^\dagger(t)\hat{q}_j\hat{U}(t)\right] \!
  \left[ \hat{U}^\dagger(t)\hat{q}_j\hat{U}(t),\hat{p}_i\right] \!
  |\Psi }.
  \label{eq:pq_OTOC_w_Ut}
\end{equation}
We take the expectation value for an initial wave packet $\Ket{\Psi}$
localized in both quadratures (like a MB coherent state,
generalizations are discussed later).

Our semiclassical method is based on approximating the path-integral
representation of $\hat{U}(t)$ in Fock space by its asymptotic form
for large $N$, the MB version~\cite{EngDujArgSchlaRicUrb13PRL} of the
Van Vleck-Gutzwiller propagator~\cite{Gut1991Book},
\begin{align}
  K(\vecfin{q},\vecinit{q};t)
  &=\Braket{\vecfin{q}| \hat{U}(t) |\vecinit{q}}
    \label{eq:VVpropagator}
  \\
  &\simeq \!\!\!\!\!\! \sum_{
    \gamma: \vecinit{q} \rightarrow \vecfin{q}
    } \!\!\!\!\!\!
    A_\gamma(\vecfin{q},\vecinit{q};t)
    \rme^{
    (\rmi / \heff)
    R_\gamma(\vecfin{q},\vecinit{q};t) }\,  .
    \nonumber
\end{align}
The sum runs over all (mean-field) solutions $\gamma$ of the classical
equations of motion
$\rmi \partial \vec{\Phi}/\partial t = \partial
\mathcal{H}^{\mathrm{cl}}/\partial \vec{\Phi}^\ast$ of the classical
Hamilton function that denotes the mean-field limit of $\hat{H}$,
Eq.~(\ref{eq:H}), for $\heff\! = \! 1/N \!\ll\! 1$:
\begin{equation}
  \mathcal{H}^{\mathrm{cl}} \! \left( \vec{q}, \vec{p} \right)
   \! = \! \frac{1}{\hbar} \sum_{ij=1}^n h_{ij}  \Phi_i^* \Phi_j
   \! + \! \frac{1}{\hbar} \sum_{ijkl=1}^n V_{ijkl}
  \Phi_i^* \Phi_j^* \Phi_k \Phi_l\, .
  \label{eq:H_cl}
\end{equation}
The initial and final real parts of the complex fields
$\vec{\Phi}\! =\! ( \vec{q} + \rmi \vec{p} ) / \sqrt{2}$ are fixed by
$\vecinit{q}$ and $\vecfin{q}$, but not their imaginary parts, thus
generally admitting many time-dependent mean-field solutions or
``trajectories'' $\gamma$ that enter the coherent sum in
Eq.~(\ref{eq:VVpropagator}) and are ultimately responsible for MB
interference effects.
In Eq.~(\ref{eq:VVpropagator}) the phases are given by classical
actions
$R_\gamma(\vecfin{q},\vecinit{q};t) \! = \!  \int_0^t \rmd t' [
\vec{p}_\gamma(t') \! \cdot \! \dot{\vec{q}}_\gamma(t') \! - \!
\mathcal{H}^{\mathrm{cl}} \! \left( \vec{q}_\gamma(t'),
  \vec{p}_\gamma(t') \right) ] $ along $\gamma$ and the weights
$A_\gamma$ reflect their classical stability [see
Eq.~(\ref{eq:stability_amplitude_squared}) in the
Supplemental Material \cite{SuppMat}].
We assume that the mean-field limit exhibits uniformly hyperbolic,
chaotic dynamics where the exponential growth has the same Lyapunov
exponent $\lambda$ at any phase space point. Here, we do not address
questions concerning light cone information spreading and nonchaotic
behavior, {\em e.g}., due to (partial) integrability or MB
localization.
Inserting unit operators in the position quadrature representation into
Eq.~(\ref{eq:pq_OTOC_w_Ut}) and using Eq.~(\ref{eq:VVpropagator}) for
$K$ we get a general semiclassical representation of the OTOC.
To leading order in $\heff$, derivatives
$\hat{p}_i\! =\! -\rmi \heff \partial / \partial q_i$ only act on the
phases in $K$ and thus, using the relations $
\partial R_{\gamma} / \partial \vecinit{q}\! =\! -\vecinit{p}_\gamma,
$ we obtain for the OTOC Eq.~(\ref{eq:pq_OTOC_w_Ut})
\begin{align}
  &C(t) \simeq
    \!\! \int\!\! \rmd^n q_1\!  \int \! \! \rmd^n q_2\!
    \int \! \! \rmd^n q_3 \! \int \! \! \rmd^n q_4\!
    \int \! \!\rmd^n q_5
    \Psi^{*}\!\left(\vec{q}_1\right)\Psi\!\left(\vec{q}_5\right)
    \nonumber \\
  & \quad  \times \!\!\!\!\!
    \sum_{
    \substack{
    \alpha': \vec{q}_1 {\rightarrow}\vec{q}_2\\
  \alpha : \vec{q}_3 {\rightarrow}\vec{q}_2}
  } \!\!\!\!
  A_{\alpha'}^*  A_{\alpha}
  \rme^{(\rmi/\heff)\left(\!-R_{\alpha'}+R_{\alpha}\right)}
  \left(\init{p}_{\alpha',i}\!-\!\init{p}_{\alpha,i} \right)\fin{q}_{\alpha,j}
  \nonumber
  \\
  & \quad \times \!\!\!\!\!
    \sum_{
    \substack{
    \beta': \vec{q}_3 {\rightarrow}\vec{q}_4\\
  \beta : \vec{q}_5 {\rightarrow}\vec{q}_4}
  }  \!\!\!\!
  A_{\beta'}^*  A_{\beta}
  \rme^{(\rmi/\heff) \left(\!-R_{\beta'}+R_{\beta}\right)} \!
  \left(\init{p}_{\beta,i}\!-\!\init{p}_{\beta',i} \right)\fin{q}_{\beta,j} \, .
  \label{eq:OTOC_sc_integral_representation}
\end{align}
The four time evolution operators in Eq.~(\ref{eq:pq_OTOC_w_Ut}) have
been transformed to fourfold sums over contributions from trajectories
of temporal length $t$ linking different initial and final position
quadratures.
A schematic illustration of a representative trajectories quadruple
that displays the geometric connections at the corresponding position
quadratures $\vec{q}_l$, $l=1,\ldots,5$, is given by
\begin{align}
   \diagram{width=0.4\linewidth}{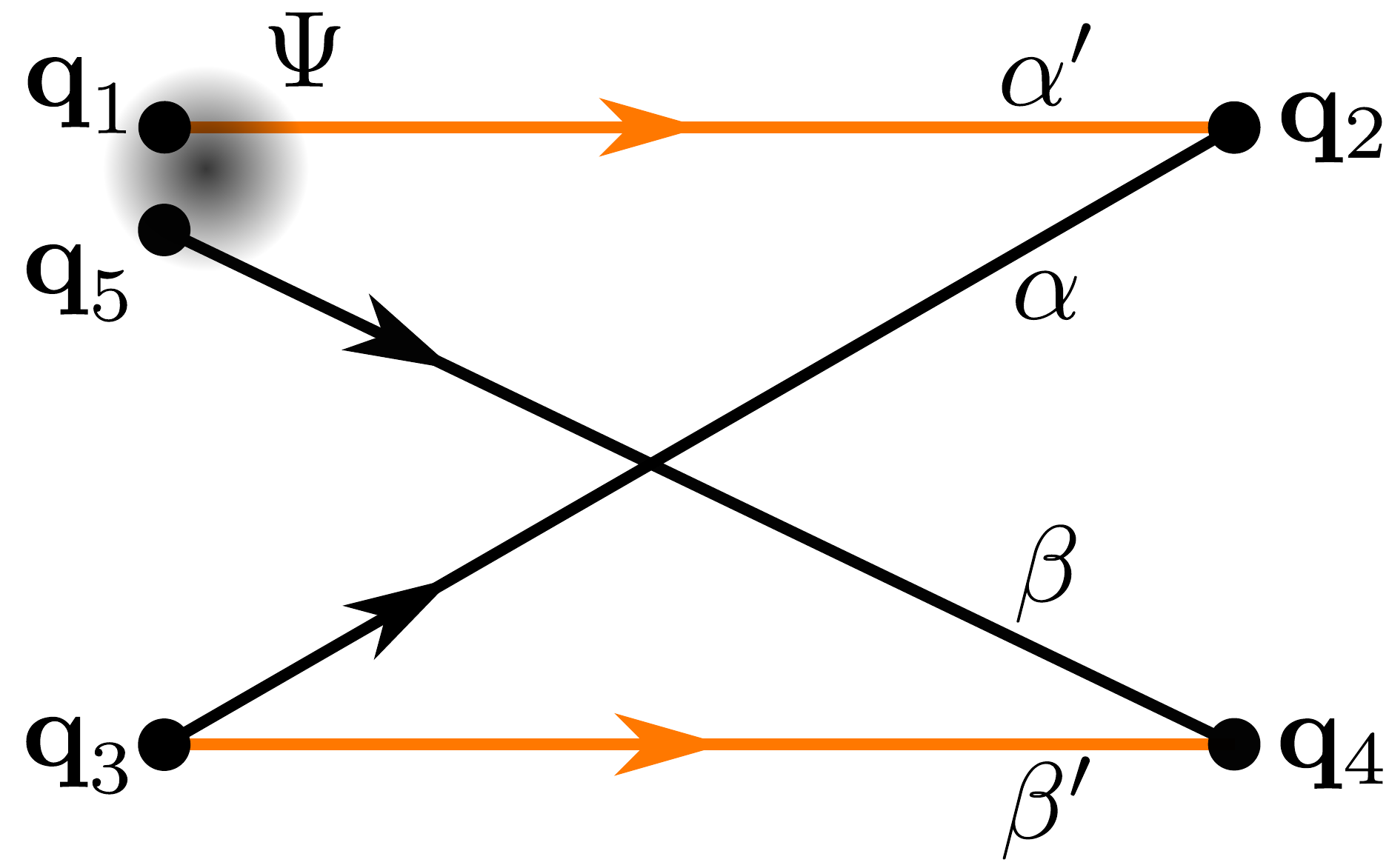} \, .
  \label{eq:OTOC_graphical_representation_sc}
\end{align}
Black (orange) arrows refer to contributions to $K$ ($K^\ast$), and
the gray shaded spot mimics the (localized) state $|\Psi\rangle$.
The semiclassical approximation in
Eq.~(\ref{eq:OTOC_sc_integral_representation}) amounts to substitute
$\hat{p}_i$, $\hat{q}_j$ in Eq.~(\ref{eq:pq_OTOC_w_Ut}) by their
classical counterparts $\init{p}_{\gamma,i}$ and $\fin{q}_{\gamma,j}$
for $\gamma\!\in\!\{\alpha,\beta,\alpha',\beta'\}$.
The commutators themselves translate into differences of initial
momenta of trajectories not restricted to start at nearby positions.

Since $R_\gamma(\vecfin{q},\vecinit{q};t) \! \gg \! \heff$ in the
semiclassical limit, the phase factors in
Eq.~(\ref{eq:OTOC_sc_integral_representation}) are generally highly
oscillatory when integrating over initial or final positions.
Hence, contributions from arbitrary trajectory quadruples are
suppressed, while correlated quadruples with action differences such
that
$R_\alpha\!-\!R_{\alpha'}\!+\!R_{\beta}\!-\!R_{\beta'} \simeq
\mathcal{O}(\heff)$ will dominantly contribute to $C(t)$.
These are constellations where most of the time trajectories are
pairwise nearly identical, except in so-called encounter regions in
phase space where trajectory pairs approach each other, follow a
correlated evolution and exchange their partners.

For OTOCs the relevant quadruples involve a single encounter and can
be subdivided into four classes depicted in
Fig.~\ref{fig:OTOC_diagram_classes}:
Diagram (a) represents a bundle of four trajectories staying in close
vicinity to each other, {\em i.e}., forming an encounter, during the
whole time $t$.
Panels (b) and (c) display ``two-leg'' diagrams with an encounter at
the beginning or end, and with uncorrelated dynamics of the two
trajectory pairs (``legs'') outside the encounter.
The ``four-leg'' diagrams in (d) are characterized by uncorrelated
motion before {\em and} after the encounter. The structure of the OTOC
implies that the two legs on the same side of an encounter are of
equal length.

Inside an encounter (boxes in Fig.~\ref{fig:OTOC_diagram_classes}) the
hyperbolic dynamics essentially follows a common mean-field solution,
{\em i.e.}, linearization around one reference trajectory allows for
expressing the remaining three trajectories.
If their action differences are of order $\heff$ the time scale
related to an encounter just corresponds to $\tauE$
[Eqs.~(\ref{eq:stable_unstable_times}), (\ref{eq:encounter_time}) and
(\ref{eq:Ehrenfest_time}) in Ref.~\cite{SuppMat}].
Because of the exponential growth of distances in chaotic phase space the
dynamics merges at the encounter boundary into uncorrelated time
evolution of two trajectory legs [see, {\em e.g.}, trajectories
$\alpha$ and $\beta$ in Fig.~\ref{fig:OTOC_diagram_classes} (b)].
Notably, Hamilton dynamics implies that this exponential separation
along unstable manifolds in phase space is complemented by motion near
stable manifolds, leading to the formation of (pairs of) \emph{exponentially
close} trajectories~\cite{SieRicPS2001}.
This mechanism gets quantum mechanically relevant for times beyond
$\tauE$ [see, {\em e.g.}, paths $\alpha'$ and $\alpha$ or $\beta$ and
$\beta'$ in Figs.~\ref{fig:OTOC_diagram_classes} (b) and (d)] and
will prove crucial for semiclassically restoring unitarity and for
explaining OTOC saturation.

The evaluation of Eq.~(\ref{eq:OTOC_sc_integral_representation})
requires a thorough consideration of the dynamics in and around the
encounter regions and the calculation of corresponding encounter
integrals based on statistical averages invoking ergodic properties of
chaotic systems.
The detailed evaluation of the diagrams (a) to (d) in
Fig.~\ref{fig:OTOC_diagram_classes} as a function of $\tauE$ for
$\heff\!\ll\! 1$ is provided in Supplemental material \cite{SuppMat}.
The $\tauE$ dependence of related objects has been considered for a
variety of spectral, scattering, and transport properties of chaotic SP
systems \cite{AdaPRB2003, BroRahPRB2006, JacWhiPRB2006,
  WalGutGouRicPRL2008, BroPRB2007,
  KuiWalPetBerRicPRL2010}. Conceptually, our derivation follows along
the lines of these works \footnote{Specific diagrams similar to class
  (d) in Fig.~\ref{fig:OTOC_diagram_classes} have been considered in
  the context of shot noise \cite{Lassl2003, BraHeuMulHaaJPA2006,
    BroPRB2007} and quantum chaotic SP \cite{KuiWalPetBerRicPRL2010}
  and MB \cite{UrbKuiHumMatRicPRL2016} scattering.}, but requires the
generalization to high-dimensional MB phase space.
Moreover, the encounter integrals involve additional amplitudes
related to the operators in the OTOC that demand special treatment,
depending on whether the initial or final position quadratures are
inside an encounter.

\begin{figure}
    \includegraphics[width=\linewidth]{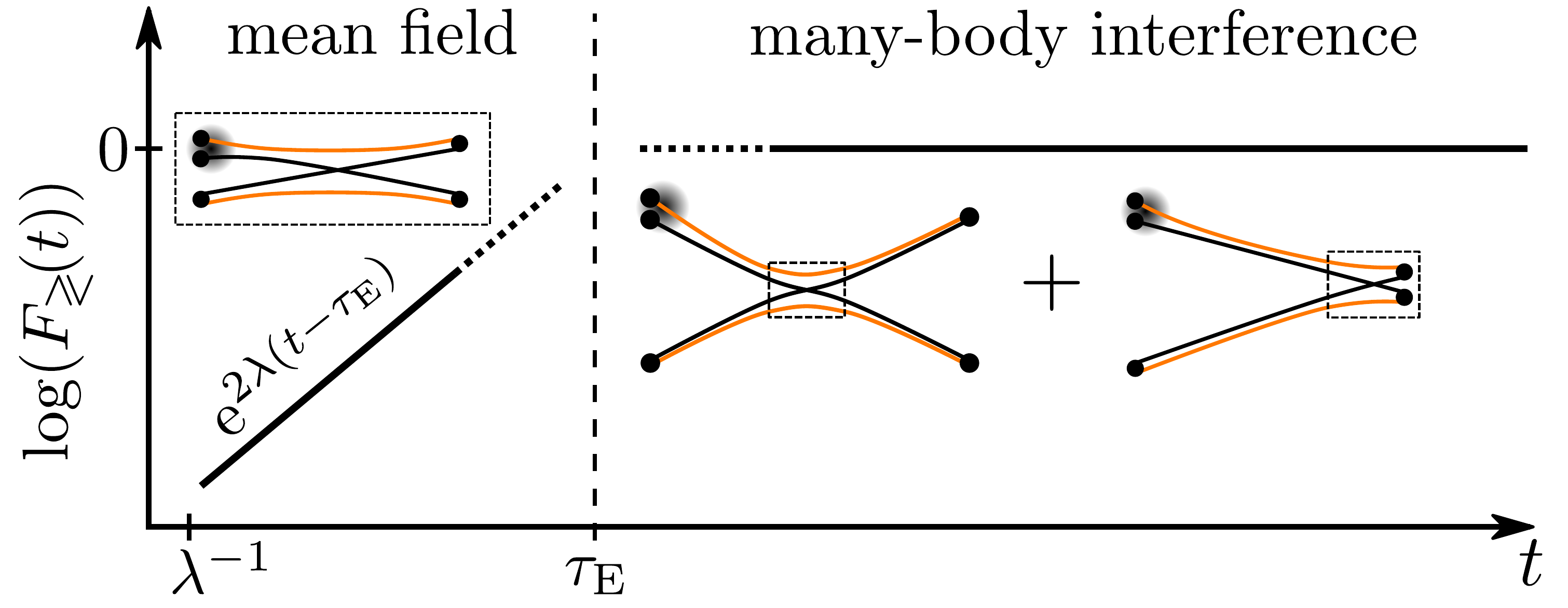}
    \caption{Universal contributions to the time evolution of the
      OTOC $C(t)$ for classically chaotic many-body quantum systems
      before [$F_<(t)$, Eq.~(\ref{eq:smaller_approx})] and after
      [$F_>(t)$, Eq.~(\ref{eq:larger_approx})] the Ehrenfest time
      $\tauE\!=\!(1/\lambda) \log(N)$ marked by the vertical dashed
      line. The insets show diagrams (a), (d), and (c) from
      Fig.~\ref{fig:OTOC_diagram_classes}, representing interfering
      mean-field solutions.  Not shown is the crossover regime at
      $t\approx \tauE$ to which all diagrams from
      Fig.~\ref{fig:OTOC_diagram_classes} contribute.}
  \label{fig:temporal_behaviour}
\end{figure}

Using furthermore the $A_\gamma$ in
Eq.~(\ref{eq:OTOC_sc_integral_representation}) to convert integrations
over final positions into initial momenta, the OTOC contribution from
each diagram is conveniently represented as phase-space average
\begin{align}
  C(t) \simeq
  \int \rmd^n q \int \rmd^n p\, W(\vec{q},\vec{p})
  I(\vec{q},\vec{p};t) \, .
  \label{eq:PS_average}
\end{align}
Here,
$ W(\vec{q},\vec{p}) \!=\! \int\!\! \rmd^n y / (2\pi\heff)^n
\Psi^*\!\left(\vec{q}\!+\! \vec{y}/2 \right)
\Psi\left(\vec{q}\!-\!\vec{y}/2 \right) $
$ \exp[(\rmi/\heff)\vec{y}\vec{p}] $ is the Wigner
function~\cite{Ozo1990} of the initial state $\Psi$, and
$I(\vec{q},\vec{p};t)$ comprises all encounter integrals.
As shown in Ref.~\cite{SuppMat} and sketched in
Fig.~\ref{fig:temporal_behaviour}, for times $t \! < \! \tauE$ the
only non-negligible contribution $I_<$ originates from diagram (a),
whereas a combination of diagrams (c) and (d) yields the contribution
$I_>$ nonvanishing for $t\! > \!\tauE$.

Using $\vecfin{x}(\vec{x};t)$ as the final phase space point of a
trajectory originating from $\vec{x}=(\vec{q},\vec{p})$, these terms
read
\begin{align}
  I_< (\vec{x};t)\!
  &=\!
    F_<\!\left(t\right) \! \left(
    \sum_{l=1}^{n-2}\!\!
    \left[\veces{l}{\vec{x}}\right]_{p_i}
    \!\!
    \left[\veceu{l}{\vecfin{x}(\vec{x};t)}\right]_{q_j}\!\!
    \right)^2\!\!\!,
    \label{eq:smaller}\\
  \!I_> (\vec{x};t)
  &=  F_>(t) \braket{
    \left(p_i-p'_i\right)^2 }_{\vec{x}}
    \left( \braket{{q'_j}^2}_{\vec{x}} - \braket{q'_j}_{\vec{x}}^2\right)  \, .
    \label{eq:larger}
\end{align}
Here  $\braket{f(\vec{x}')}_{\vec{x}}$ denotes the
average of a phase-space function $f$ over the manifold defined
through $\vec{x}$ by constant energy and particle density
[Eq.~(\ref{eq:def_ergodic_PS_average}) in Ref.~\cite{SuppMat}].
In Eq.~(\ref{eq:smaller}) the vectors $\vec{e}_{s/u}^{(l)}(\vec{x})$
denote the $n\!-\!2$ directions towards the stable, respectively
unstable manifolds at $\vec{x}$, and the labels $q_j$, $p_i$ indicate
components of those.
Finally, in Eqs.~(\ref{eq:smaller}, \ref{eq:larger})
\begin{align}
  F_<(t)
  & =
    \rme^{2\lambda(t-\tauE)}\left(\frac{2}{\pi}\right)^{n-2}
    \left[\Si\left(\rme^{\lambda(\tauE-t)}\right)\right]^{n-4}\nonumber\\
  &\quad \times
    \left[\Si\left(\rme^{\lambda(\tauE-t)}\right)
    -\sin\left(\rme^{\lambda(\tauE-t)}\right)\right]^2
    \,,
  \\
  F_>(t)
  & =  \left[
    \frac{2}{\pi} \Si\left( \rme^{\lambda \tauE} \right) \right]^{n-2}
    \hspace{-2ex}-
    \left[ \frac{2}{\pi} \Si\left(\rme^{\lambda(\tauE-t)}\right) \right]^{n-2}
    \label{eq:temporal_behaviour_larger}
\end{align}
with $\Si(z)\!=\!\int_0^z (\sin (z')/z') \rmd z'$. In the semiclassical
limit follows $\lambda\tauE \! =\! \log(1/\heff)\!\gg\! 1$ such that
$F_<(t\!>\!\tauE)$ is strongly suppressed (reflecting the vanishing phase
space volume of quadruples of trajectories remaining close to each
other over longer times) and can be expressed by a Heaviside
step function,
\begin{equation}
  F_<(t)\approx\rme^{2\lambda(t-\tauE)}\Theta(\tauE-t)
  =\heff^2\rme^{2\lambda t} \Theta(\tauE-t)\,.
  \label{eq:smaller_approx}
\end{equation}
As a result the contribution to $C(t)$ in Eq.~(\ref{eq:PS_average}),
associated with $I_<$ and $F_<(t)$, is responsible for the initial
exponential growth $\exp[2\lambda(t-\tauE)]$ of the OTOC for
$t\!<\!\tauE$, as also depicted in
Fig.~\ref{fig:temporal_behaviour}. It reflects unstable mean-field
behavior.
Note that for $t\! >\! \lambda^{-1}$ (the ergodic time) \cite{Gas1998Book}
\begin{equation}
    \frac{\partial \fin{q}_j}{\partial \init{q}_i}(\vec{x};t)
   \approx\!
   \sum_{l=1}^{n-2}
   \left[
     \veces{l}{\vec{x}}
   \right]_{p_i}\!
   \left[
     \veceu{l}{\vecfin{x}(\vec{x};t)}
   \right]_{q_j}\!
   \rme^{\lambda t}\,  ,
\end{equation}
implying that our result, Eq.~(\ref{eq:smaller}), reduces to the short-time
limit, Eq.~(\ref{eq:OTOC_Moyal}), of the commutator, but moreover
additionally contains the missing cutoff through $\theta(\tauE-t)$.

On the contrary, $F_>(t)$ in Eq.~(\ref{eq:temporal_behaviour_larger})
is suppressed for $t\!<\!\tauE$, but is indeed responsible for
post-Ehrenfest OTOC saturation, as for $\lambda \tauE\! \gg\! 1$ it
can be approximated by
\begin{equation}
  F_>(t)\approx\Theta(t-\tauE)\,.
  \label{eq:larger_approx}
\end{equation}
The underlying diagrams (c) and (d) represent dynamics swapping forth
and back along distinctly different encounter-coupled mean-field
solutions.
This mechanism that emerges evidently in a regime where mean-field
approaches fail~\cite{HanWuPRA2016} creates quantum correlations and
entanglement, respectively\footnote{It may be viewed as the
  underlying dynamical mechanism, supporting (in the large-$N$ limit)
  models for OTOCs based on coupled
  binaries~\cite{RakPolKey2018PRX}.}.
The underlying MB interference, accounted for in the encounter
integrals, is at the heart of $F_>(t) $ entering $I_> (\vec{x};t)$ in
Eq.~(\ref{eq:larger}).

The latter further contains classical quantities that determine its
saturation value:
the variance of the $j$th final position quadrature
$ (\Delta q'_j)^2\!=\!\braket{{q'_j}^2}_{\vec{x}} \!-\!
\braket{{q'_j}}_{\vec{x}}^2 $ and
$\braket{(p_i\!-\!p'_i)^2}_{\vec{x}}$.
A straightforward calculation of the ergodic averages, exploiting the
connection between ${p'_i}^2$ and ${q'_j}^2$ with the particle density
[see Eq.~(\ref{eq:particle_density}) in Ref.~\cite{SuppMat}] yields
$ I_> (\vec{x};t) \approx \theta(t-\tauE) (p_i^2 + 1/n )\!  \times\!
(1/n) $.

For an initial state $\ket{\Psi}$ with a Wigner function sharply
localized in phase space, the average Eq.~(\ref{eq:PS_average}) then
gives
\begin{equation}
  C(t)\approx \frac{2}{n^2}  \textrm{ for } t>\tauE\,,
  \label{eq:universal_value}
\end{equation}
with corrections of $\mathcal{O}(\heff)$ due to the finite
width.
Interestingly, the same result, Eq.~(\ref{eq:universal_value}), holds
if $\ket{\Psi}$ is an extended chaotic MB state with fixed energy and
particle density.

We finally discuss several implications and conclusions:

(i) \emph{Generalization to OTOCs with other operators.--}The entire
line of reasoning can be generalized to OTOCs involving
 operators that are smooth functions of
position and momentum quadratures for which a corresponding classical
symbol exists~\cite{SuppMat}.

(ii) {\em Time-reversal (TR) invariance and higher-order quantum
corrections.--}Remarkably, the leading quantum correction
[Fig.~\ref{fig:OTOC_diagram_classes}(d)] is of the same order as the
classical mean-field contribution at $\tauE$.
Moreover, the absence of trajectory loops in the diagrams in
Fig.~\ref{fig:OTOC_diagram_classes}, usually associated with weak
localization-like corrections, implies that our results hold true for
systems with and without TR symmetry.
Diagrams involving more than one trajectory encounter generally yield
further subleading contributions that can be susceptible to TR
symmetry breaking.
Their evaluation for OTOCs requires further research.

(iii) {\em Small-$\hbar$ limit and SP systems.--}Our semiclassical
calculation of OTOCs in the large-$N$ limit can be readily generalized
to systems of $N$ particles in $d$ spatial dimensions in the
complementary limit of small $\hbar$, including the quantum chaotic SP
case $N\!=\! 1$.
There, $\heff\!=\!\hbar/S \!\sim\! \lambda_{\rm dB}/L$ where
 $\lambda_{\mathrm{dB}}$ is the de Broglie wavelength, and $S$ and
 $L$ are typical actions and length scales of the chaotic classical limit.
Invoking the Gutzwiller propagator~\cite{Gut1991Book} in $n=d\cdot N$
dimensions in Eq.~(\ref{eq:VVpropagator})
the exponential increase of the OTOC $C_N(t)$ is then determined by
the leading Lyapunov exponent $\lambda_N$ of the corresponding
classical $N$-particle system (see, {\em e.g.}, Refs.~\cite{Kur2018JSP,
RoGaGaPRL2017} for $N\!=\!1$).
Our derivation shows that saturation sets in at the
corresponding Ehrenfest time $ \tauE^{(N)} \! \sim \!  (1/\lambda_N)
\log (\heff^{-1})$.
We can again evaluate $C_{N}(t)$ for
$t>\tauE^{(N)}$. For example, for chaotic billiards
$\braket{(p_i-p_i')^2}\!=\!p_i^2 \!+\! p^2/n$.
Since $L$ corresponds to the overall system size $\cal{L}$, $(\Delta
q'_j)^2\! \propto \! (\mathcal{L})^2\! =\! L^2$. Thus
$C_{N}(t)\!\propto\!  S^2 / n $, where the typical action $S\!=\!
\hbar/\heff$ arises here since
$[\hat{q}_j,\hat{p}_i]\!=\!\rmi\delta_{ij}\hbar
\!=\!\rmi\delta_{ij}S\heff$. Within this line of reasoning, one can
view Ref.~\cite{RoGaGaPRL2017} as a quantitative numerical
confirmation of our semiclassical findings.

Interestingly, for many systems we can have $L \ll \cal{L}$, such
as for the famous Lorentz gas~\cite{Gas1998Book}.
It is composed of scattering disks or spheres for $d\!=\!2$ or
3~\footnote{Ehrenfest time effects in Lorentz gases were studied, {\it
e.g.}, in Refs.~\cite{AleLarPRB1996, YevLutWeiRicPRL2000, BroPRB2007}.}
with diameters setting the scale $L$.
Then the dynamics is hyperbolic up to $\tauE^{(1)}$ before it becomes
diffusive.
This implies that $(\Delta q'_j)^2$ in Eq.~(\ref{eq:larger}) scales
linearly with time, $(\Delta q'_j)^2 \sim D t $, with diffusion
constant $D$.
Thus, beyond $\tau_E^{(1)}$ we expect $C_1(t)$ to first linearly
increase before it saturates at the ergodic (Thouless) time
${\cal{L}}^2 / D$.
In SP systems with diffusive dynamics arising from quantum scattering
at impurities, the transport time $t_{\mathrm{tr}}$ takes the role of
$\tau_E^{(1)}$.
This implies a sharp increase of $C_1(t)$ for $t<t_{\mathrm{tr}}$, as
already predicted in Ref.~\cite{LarOvcJETP1969}, followed by the
diffusive behavior discussed above.

(iv) {\em Nonergodic many-body dynamics.--}The nonlinear mean-field
dynamics associated with the classical limit of MB Fock space is much
less understood \cite{BorIzrSanarxiv2018, TomSchlaUllUrbRic2018PRA,
Tom2018PRE} than its SP counterpart.
If the MB dynamics is diffusive for $t\!>\!\tauE$, we expect a similar
time dependence for $C(t)$ as discussed in (iii).
The propagator Eq.~(\ref{eq:VVpropagator}) is not restricted to chaotic
dynamics, but also allows for investigating the imprint of more
complex, {\it e.g.}, mixed regular-chaotic, phase space dynamics on
OTOCs or, more generally, on the stability of MB quantum evolution
{\em per se}.

To conclude, we considered the time evolution of OTOCs by developing a
general semiclassical approach for interacting large-$N$ systems.
It links chaotic motion in the classical mean-field limit to the
correlated quantum many-body dynamics in terms of interference between
mean-field solutions giving rise to scrambling and entanglement.
We uncovered the relevant many-body quantum interference mechanism
that is responsible for the commonly observed saturation of OTOCs at
the scrambling or Ehrenfest time.
While we explicitly derived OTOCs for bosonic systems, similar
considerations should be possible for fermionic
many-body systems \footnote{To this end, the semiclassical (large-$N$)
approximation for the microscopic path integral propagator of discrete
fermionic quantum fields~\cite{EngPloUrbRic2014TCA} can be
employed.
Based on this fermionic propagator, a semiclassical calculation of a
MB spin echo gave perfect agreement with numerical quantum
calculations, see  Ref.~\cite{EngUrbRicSchla2014PRA}.
}
posing an interesting problem for future research.

\acknowledgments We thank T.~Engl, B.~Geiger, S.~Tomsovic, D.~Ullmo,
and D.~Waltner for helpful conversations. We acknowledge funding from
Deutsche Forschungsgemeinschaft through project Ri681/14-1.

\bibliographystyle{apsrev4-1}
\bibliography{ms}

\begin{thebibliography}{65}%
\makeatletter
\providecommand \@ifxundefined [1]{%
 \@ifx{#1\undefined}
}%
\providecommand \@ifnum [1]{%
 \ifnum #1\expandafter \@firstoftwo
 \else \expandafter \@secondoftwo
 \fi
}%
\providecommand \@ifx [1]{%
 \ifx #1\expandafter \@firstoftwo
 \else \expandafter \@secondoftwo
 \fi
}%
\providecommand \natexlab [1]{#1}%
\providecommand \enquote  [1]{``#1''}%
\providecommand \bibnamefont  [1]{#1}%
\providecommand \bibfnamefont [1]{#1}%
\providecommand \citenamefont [1]{#1}%
\providecommand \href@noop [0]{\@secondoftwo}%
\providecommand \href [0]{\begingroup \@sanitize@url \@href}%
\providecommand \@href[1]{\@@startlink{#1}\@@href}%
\providecommand \@@href[1]{\endgroup#1\@@endlink}%
\providecommand \@sanitize@url [0]{\catcode `\\12\catcode `\$12\catcode
  `\&12\catcode `\#12\catcode `\^12\catcode `\_12\catcode `\%12\relax}%
\providecommand \@@startlink[1]{}%
\providecommand \@@endlink[0]{}%
\providecommand \url  [0]{\begingroup\@sanitize@url \@url }%
\providecommand \@url [1]{\endgroup\@href {#1}{\urlprefix }}%
\providecommand \urlprefix  [0]{URL }%
\providecommand \Eprint [0]{\href }%
\providecommand \doibase [0]{http://dx.doi.org/}%
\providecommand \selectlanguage [0]{\@gobble}%
\providecommand \bibinfo  [0]{\@secondoftwo}%
\providecommand \bibfield  [0]{\@secondoftwo}%
\providecommand \translation [1]{[#1]}%
\providecommand \BibitemOpen [0]{}%
\providecommand \bibitemStop [0]{}%
\providecommand \bibitemNoStop [0]{.\EOS\space}%
\providecommand \EOS [0]{\spacefactor3000\relax}%
\providecommand \BibitemShut  [1]{\csname bibitem#1\endcsname}%
\let\auto@bib@innerbib\@empty
\bibitem [{\citenamefont {Gutzwiller}(1991)}]{Gut1991Book}%
  \BibitemOpen
  \bibfield  {author} {\bibinfo {author} {\bibfnamefont {M.~C.}\ \bibnamefont
  {Gutzwiller}},\ }\href {https://www.springer.com/de/book/9780387971735}
  {\emph {\bibinfo {title} {Chaos in Classical and Quantum Mechanics}}}\
  (\bibinfo  {publisher} {Springer New York},\ \bibinfo {year}
  {1991})\BibitemShut {NoStop}%
\bibitem [{Note1()}]{Note1}%
  \BibitemOpen
  \bibinfo {note} {A. Kitaev, \protect \emph {Hidden Correlations in the
  Hawking Radiation and Thermal Noise}, talk at Breakthrough Physics Prize
  Symposium, Nov. 10, 2014, \protect \url
  {https://www.youtube.com/watch?v=OQ9qN8j7EZI}}\BibitemShut {NoStop}%
\bibitem [{\citenamefont {Sekino}\ and\ \citenamefont
  {Susskind}(2008)}]{SekSusJHEP2008}%
  \BibitemOpen
  \bibfield  {author} {\bibinfo {author} {\bibfnamefont {Y.}~\bibnamefont
  {Sekino}}\ and\ \bibinfo {author} {\bibfnamefont {L.}~\bibnamefont
  {Susskind}},\ }\href {\doibase 10.1088/1126-6708/2008/10/065} {\bibfield
  {journal} {\bibinfo  {journal} {J.~High Energ.~Phys.}\ }\textbf {\bibinfo
  {volume} {10}},\ \bibinfo {pages} {65} (\bibinfo {year} {2008})}\BibitemShut
  {NoStop}%
\bibitem [{\citenamefont {Shenker}\ and\ \citenamefont
  {Stanford}(2014)}]{SheStaJHEP2014}%
  \BibitemOpen
  \bibfield  {author} {\bibinfo {author} {\bibfnamefont {S.~H.}\ \bibnamefont
  {Shenker}}\ and\ \bibinfo {author} {\bibfnamefont {D.}~\bibnamefont
  {Stanford}},\ }\href {\doibase 10.1007/JHEP03(2014)067} {\bibfield  {journal}
  {\bibinfo  {journal} {J.~High Energ.~Phys.}\ }\textbf {\bibinfo {volume}
  {2014}},\ \bibinfo {pages} {67} (\bibinfo {year} {2014})}\BibitemShut
  {NoStop}%
\bibitem [{\citenamefont {Maldacena}\ \emph {et~al.}(2016)\citenamefont
  {Maldacena}, \citenamefont {Shenker},\ and\ \citenamefont
  {Stanford}}]{MalSheSta2015}%
  \BibitemOpen
  \bibfield  {author} {\bibinfo {author} {\bibfnamefont {J.}~\bibnamefont
  {Maldacena}}, \bibinfo {author} {\bibfnamefont {S.~H.}\ \bibnamefont
  {Shenker}}, \ and\ \bibinfo {author} {\bibfnamefont {D.}~\bibnamefont
  {Stanford}},\ }\href {\doibase 10.1007/JHEP08(2016)106} {\bibfield  {journal}
  {\bibinfo  {journal} {J.~High Energ.~Phys.}\ }\textbf {\bibinfo {volume}
  {2016}},\ \bibinfo {pages} {106} (\bibinfo {year} {2016})}\BibitemShut
  {NoStop}%
\bibitem [{\citenamefont {Larkin}\ and\ \citenamefont
  {Ovchinnikov}(1969)}]{LarOvcJETP1969}%
  \BibitemOpen
  \bibfield  {author} {\bibinfo {author} {\bibfnamefont {A.~I.}\ \bibnamefont
  {Larkin}}\ and\ \bibinfo {author} {\bibfnamefont {Y.~N.}\ \bibnamefont
  {Ovchinnikov}},\ }\href
  {http://www.jetp.ac.ru/cgi-bin/e/index/e/28/6/p1200?a=list} {\bibfield
  {journal} {\bibinfo  {journal} {Soviet Phys.~JETP}\ }\textbf {\bibinfo
  {volume} {28}},\ \bibinfo {pages} {1200} (\bibinfo {year}
  {1969})}\BibitemShut {NoStop}%
\bibitem [{\citenamefont {Zhu}\ \emph {et~al.}(2016)\citenamefont {Zhu},
  \citenamefont {Hafezi},\ and\ \citenamefont {Grover}}]{ZhuHafGroPRA2016}%
  \BibitemOpen
  \bibfield  {author} {\bibinfo {author} {\bibfnamefont {G.}~\bibnamefont
  {Zhu}}, \bibinfo {author} {\bibfnamefont {M.}~\bibnamefont {Hafezi}}, \ and\
  \bibinfo {author} {\bibfnamefont {T.}~\bibnamefont {Grover}},\ }\href
  {\doibase 10.1103/PhysRevA.94.062329} {\bibfield  {journal} {\bibinfo
  {journal} {Phys.~Rev.~A}\ }\textbf {\bibinfo {volume} {94}},\ \bibinfo
  {pages} {062329} (\bibinfo {year} {2016})}\BibitemShut {NoStop}%
\bibitem [{\citenamefont {Swingle}\ \emph {et~al.}(2016)\citenamefont
  {Swingle}, \citenamefont {Bentsen}, \citenamefont {Schleier-Smith},\ and\
  \citenamefont {Hayden}}]{SwiBenSchleHazPRA2016}%
  \BibitemOpen
  \bibfield  {author} {\bibinfo {author} {\bibfnamefont {B.}~\bibnamefont
  {Swingle}}, \bibinfo {author} {\bibfnamefont {G.}~\bibnamefont {Bentsen}},
  \bibinfo {author} {\bibfnamefont {M.}~\bibnamefont {Schleier-Smith}}, \ and\
  \bibinfo {author} {\bibfnamefont {P.}~\bibnamefont {Hayden}},\ }\href
  {\doibase 10.1103/PhysRevA.94.040302} {\bibfield  {journal} {\bibinfo
  {journal} {Phys.~Rev.~A}\ }\textbf {\bibinfo {volume} {94}},\ \bibinfo
  {pages} {040302} (\bibinfo {year} {2016})}\BibitemShut {NoStop}%
\bibitem [{\citenamefont {Campisi}\ and\ \citenamefont
  {Goold}(2017)}]{CamGooPRE2016}%
  \BibitemOpen
  \bibfield  {author} {\bibinfo {author} {\bibfnamefont {M.}~\bibnamefont
  {Campisi}}\ and\ \bibinfo {author} {\bibfnamefont {J.}~\bibnamefont
  {Goold}},\ }\href {\doibase 10.1103/PhysRevE.95.062127} {\bibfield  {journal}
  {\bibinfo  {journal} {Phys.~Rev.~E}\ }\textbf {\bibinfo {volume} {95}},\
  \bibinfo {pages} {062127} (\bibinfo {year} {2017})}\BibitemShut {NoStop}%
\bibitem [{\citenamefont {Li}\ \emph {et~al.}(2017)\citenamefont {Li},
  \citenamefont {Fan}, \citenamefont {Wang}, \citenamefont {Ye}, \citenamefont
  {Zeng}, \citenamefont {Zhai}, \citenamefont {Peng},\ and\ \citenamefont
  {Du}}]{LiFanWanZeYenYhaPenDuPRX2016}%
  \BibitemOpen
  \bibfield  {author} {\bibinfo {author} {\bibfnamefont {J.}~\bibnamefont
  {Li}}, \bibinfo {author} {\bibfnamefont {R.}~\bibnamefont {Fan}}, \bibinfo
  {author} {\bibfnamefont {H.}~\bibnamefont {Wang}}, \bibinfo {author}
  {\bibfnamefont {B.}~\bibnamefont {Ye}}, \bibinfo {author} {\bibfnamefont
  {B.}~\bibnamefont {Zeng}}, \bibinfo {author} {\bibfnamefont {H.}~\bibnamefont
  {Zhai}}, \bibinfo {author} {\bibfnamefont {X.}~\bibnamefont {Peng}}, \ and\
  \bibinfo {author} {\bibfnamefont {J.}~\bibnamefont {Du}},\ }\href {\doibase
  10.1103/PhysRevX.7.031011} {\bibfield  {journal} {\bibinfo  {journal}
  {Phys.~Rev.~X}\ }\textbf {\bibinfo {volume} {7}},\ \bibinfo {pages} {031011}
  (\bibinfo {year} {2017})}\BibitemShut {NoStop}%
\bibitem [{\citenamefont {G{\"a}rttner}\ \emph {et~al.}(2017)\citenamefont
  {G{\"a}rttner}, \citenamefont {Bohnet}, \citenamefont {Safavi-Naini},
  \citenamefont {Wall}, \citenamefont {Bollinger},\ and\ \citenamefont
  {Rey}}]{GarBohSafWalBolRey2016}%
  \BibitemOpen
  \bibfield  {author} {\bibinfo {author} {\bibfnamefont {M.}~\bibnamefont
  {G{\"a}rttner}}, \bibinfo {author} {\bibfnamefont {J.~G.}\ \bibnamefont
  {Bohnet}}, \bibinfo {author} {\bibfnamefont {A.}~\bibnamefont
  {Safavi-Naini}}, \bibinfo {author} {\bibfnamefont {M.~L.}\ \bibnamefont
  {Wall}}, \bibinfo {author} {\bibfnamefont {J.~J.}\ \bibnamefont {Bollinger}},
  \ and\ \bibinfo {author} {\bibfnamefont {A.~M.}\ \bibnamefont {Rey}},\ }\href
  {\doibase 10.1038/NPHYS4119} {\bibfield  {journal} {\bibinfo  {journal}
  {Nat.~Phys.}\ }\textbf {\bibinfo {volume} {13}},\ \bibinfo {pages} {781}
  (\bibinfo {year} {2017})}\BibitemShut {NoStop}%
\bibitem [{\citenamefont {Kurchan}(2018)}]{Kur2018JSP}%
  \BibitemOpen
  \bibfield  {author} {\bibinfo {author} {\bibfnamefont {J.}~\bibnamefont
  {Kurchan}},\ }\href {\doibase 10.1007/s10955-018-2052-7} {\bibfield
  {journal} {\bibinfo  {journal} {J.~Stat.~Phys.}\ }\textbf {\bibinfo {volume}
  {171}},\ \bibinfo {pages} {965} (\bibinfo {year} {2018})}\BibitemShut
  {NoStop}%
\bibitem [{\citenamefont {Jalabert}\ and\ \citenamefont
  {Pastawski}(2001)}]{JalPasPRL2001}%
  \BibitemOpen
  \bibfield  {author} {\bibinfo {author} {\bibfnamefont {R.~A.}\ \bibnamefont
  {Jalabert}}\ and\ \bibinfo {author} {\bibfnamefont {H.~M.}\ \bibnamefont
  {Pastawski}},\ }\href {\doibase 10.1103/PhysRevLett.86.2490} {\bibfield
  {journal} {\bibinfo  {journal} {Phys.~Rev.~Lett.}\ }\textbf {\bibinfo
  {volume} {86}},\ \bibinfo {pages} {2490} (\bibinfo {year}
  {2001})}\BibitemShut {NoStop}%
\bibitem [{\citenamefont {Rozenbaum}\ \emph {et~al.}(2017)\citenamefont
  {Rozenbaum}, \citenamefont {Ganeshan},\ and\ \citenamefont
  {Galitski}}]{RoGaGaPRL2017}%
  \BibitemOpen
  \bibfield  {author} {\bibinfo {author} {\bibfnamefont {E.~B.}\ \bibnamefont
  {Rozenbaum}}, \bibinfo {author} {\bibfnamefont {S.}~\bibnamefont {Ganeshan}},
  \ and\ \bibinfo {author} {\bibfnamefont {V.}~\bibnamefont {Galitski}},\
  }\href {\doibase 10.1103/PhysRevLett.118.086801} {\bibfield  {journal}
  {\bibinfo  {journal} {Phys.~Rev.~Lett.}\ }\textbf {\bibinfo {volume} {118}},\
  \bibinfo {pages} {086801} (\bibinfo {year} {2017})}\BibitemShut {NoStop}%
\bibitem [{\citenamefont {Bagrets}\ \emph {et~al.}(2017)\citenamefont
  {Bagrets}, \citenamefont {Altland},\ and\ \citenamefont
  {Kamenev}}]{BagAltKamNPB2017}%
  \BibitemOpen
  \bibfield  {author} {\bibinfo {author} {\bibfnamefont {D.}~\bibnamefont
  {Bagrets}}, \bibinfo {author} {\bibfnamefont {A.}~\bibnamefont {Altland}}, \
  and\ \bibinfo {author} {\bibfnamefont {A.}~\bibnamefont {Kamenev}},\ }\href
  {\doibase 10.1016/j.nuclphysb.2017.06.012} {\bibfield  {journal} {\bibinfo
  {journal} {Nuclear Phys.~B}\ }\textbf {\bibinfo {volume} {921}},\ \bibinfo
  {pages} {727} (\bibinfo {year} {2017})}\BibitemShut {NoStop}%
\bibitem [{\citenamefont {Scaffidi}\ and\ \citenamefont
  {Altman}(2017)}]{ScaAltarxiv2017}%
  \BibitemOpen
  \bibfield  {author} {\bibinfo {author} {\bibfnamefont {T.}~\bibnamefont
  {Scaffidi}}\ and\ \bibinfo {author} {\bibfnamefont {E.}~\bibnamefont
  {Altman}},\ }\href@noop {} {} (\bibinfo {year} {2017}),\ \Eprint
  {http://arxiv.org/abs/1711.04768} {arXiv:1711.04768} \BibitemShut {NoStop}%
\bibitem [{\citenamefont {Torres-Herrera}\ \emph {et~al.}(2018)\citenamefont
  {Torres-Herrera}, \citenamefont {Garc\'{\i}a-Garc\'{\i}a},\ and\
  \citenamefont {Santos}}]{TorGarSan2018PRB}%
  \BibitemOpen
  \bibfield  {author} {\bibinfo {author} {\bibfnamefont {E.~J.}\ \bibnamefont
  {Torres-Herrera}}, \bibinfo {author} {\bibfnamefont {A.~M.}\ \bibnamefont
  {Garc\'{\i}a-Garc\'{\i}a}}, \ and\ \bibinfo {author} {\bibfnamefont {L.~F.}\
  \bibnamefont {Santos}},\ }\href {\doibase 10.1103/PhysRevB.97.060303}
  {\bibfield  {journal} {\bibinfo  {journal} {Phys.~Rev.~B}\ }\textbf {\bibinfo
  {volume} {97}},\ \bibinfo {pages} {060303} (\bibinfo {year}
  {2018})}\BibitemShut {NoStop}%
\bibitem [{\citenamefont {Cotler}\ \emph {et~al.}(2017)\citenamefont {Cotler},
  \citenamefont {Hunter-Jones}, \citenamefont {Liu},\ and\ \citenamefont
  {Yoshida}}]{CotHunKLiuZosJHEP2017}%
  \BibitemOpen
  \bibfield  {author} {\bibinfo {author} {\bibfnamefont {J.}~\bibnamefont
  {Cotler}}, \bibinfo {author} {\bibfnamefont {N.}~\bibnamefont
  {Hunter-Jones}}, \bibinfo {author} {\bibfnamefont {J.}~\bibnamefont {Liu}}, \
  and\ \bibinfo {author} {\bibfnamefont {B.}~\bibnamefont {Yoshida}},\ }\href
  {\doibase 10.1007/JHEP11(2017)048} {\bibfield  {journal} {\bibinfo  {journal}
  {J.~High Energ.~Phys.}\ }\textbf {\bibinfo {volume} {2017}},\ \bibinfo
  {pages} {48} (\bibinfo {year} {2017})}\BibitemShut {NoStop}%
\bibitem [{\citenamefont {{del Campo}}\ \emph {et~al.}(2017)\citenamefont {{del
  Campo}}, \citenamefont {Molina-Vilaplana},\ and\ \citenamefont
  {Sonner}}]{DelMolSonPRD2017}%
  \BibitemOpen
  \bibfield  {author} {\bibinfo {author} {\bibfnamefont {A.}~\bibnamefont {{del
  Campo}}}, \bibinfo {author} {\bibfnamefont {J.}~\bibnamefont
  {Molina-Vilaplana}}, \ and\ \bibinfo {author} {\bibfnamefont
  {J.}~\bibnamefont {Sonner}},\ }\href {\doibase 10.1103/PhysRevD.95.126008}
  {\bibfield  {journal} {\bibinfo  {journal} {Phys.~Rev.~D}\ }\textbf {\bibinfo
  {volume} {95}},\ \bibinfo {pages} {126008} (\bibinfo {year}
  {2017})}\BibitemShut {NoStop}%
\bibitem [{\citenamefont {Bohrdt}\ \emph {et~al.}(2017)\citenamefont {Bohrdt},
  \citenamefont {Mendl}, \citenamefont {Endres},\ and\ \citenamefont
  {Knap}}]{BohMenEndKnaNJP2016}%
  \BibitemOpen
  \bibfield  {author} {\bibinfo {author} {\bibfnamefont {A.}~\bibnamefont
  {Bohrdt}}, \bibinfo {author} {\bibfnamefont {C.~B.}\ \bibnamefont {Mendl}},
  \bibinfo {author} {\bibfnamefont {M.}~\bibnamefont {Endres}}, \ and\ \bibinfo
  {author} {\bibfnamefont {M.}~\bibnamefont {Knap}},\ }\href {\doibase
  10.1088/1367-2630/aa719b} {\bibfield  {journal} {\bibinfo  {journal} {New
  J.~Phys.}\ }\textbf {\bibinfo {volume} {19}},\ \bibinfo {pages} {063001}
  (\bibinfo {year} {2017})}\BibitemShut {NoStop}%
\bibitem [{\citenamefont {Shen}\ \emph {et~al.}(2017)\citenamefont {Shen},
  \citenamefont {Zhang}, \citenamefont {Fan},\ and\ \citenamefont
  {Zhai}}]{SheZhaFanZhaPRB2016}%
  \BibitemOpen
  \bibfield  {author} {\bibinfo {author} {\bibfnamefont {H.}~\bibnamefont
  {Shen}}, \bibinfo {author} {\bibfnamefont {P.}~\bibnamefont {Zhang}},
  \bibinfo {author} {\bibfnamefont {R.}~\bibnamefont {Fan}}, \ and\ \bibinfo
  {author} {\bibfnamefont {H.}~\bibnamefont {Zhai}},\ }\href {\doibase
  10.1103/PhysRevB.96.054503} {\bibfield  {journal} {\bibinfo  {journal}
  {Phys.~Rev.~B}\ }\textbf {\bibinfo {volume} {96}},\ \bibinfo {pages} {054503}
  (\bibinfo {year} {2017})}\BibitemShut {NoStop}%
\bibitem [{\citenamefont {Hashimoto}\ \emph {et~al.}(2017)\citenamefont
  {Hashimoto}, \citenamefont {Murata},\ and\ \citenamefont
  {Yoshii}}]{HasMurYosJHEP2017}%
  \BibitemOpen
  \bibfield  {author} {\bibinfo {author} {\bibfnamefont {K.}~\bibnamefont
  {Hashimoto}}, \bibinfo {author} {\bibfnamefont {K.}~\bibnamefont {Murata}}, \
  and\ \bibinfo {author} {\bibfnamefont {R.}~\bibnamefont {Yoshii}},\ }\href
  {\doibase 10.1007/JHEP10(2017)138} {\bibfield  {journal} {\bibinfo  {journal}
  {J.~High Energ.~Phys.}\ }\textbf {\bibinfo {volume} {2017}},\ \bibinfo
  {pages} {138} (\bibinfo {year} {2017})}\BibitemShut {NoStop}%
\bibitem [{\citenamefont {Ehrenfest}(1927)}]{EhrZfP1927}%
  \BibitemOpen
  \bibfield  {author} {\bibinfo {author} {\bibfnamefont {P.}~\bibnamefont
  {Ehrenfest}},\ }\href {\doibase 10.1007/BF01329203} {\bibfield  {journal}
  {\bibinfo  {journal} {Zeit.~{f\"ur} Phys.}\ }\textbf {\bibinfo {volume}
  {45}},\ \bibinfo {pages} {455} (\bibinfo {year} {1927})}\BibitemShut
  {NoStop}%
\bibitem [{\citenamefont {Berman}\ and\ \citenamefont
  {Zaslavsky}(1978)}]{BerZasPA1978}%
  \BibitemOpen
  \bibfield  {author} {\bibinfo {author} {\bibfnamefont {G.~P.}\ \bibnamefont
  {Berman}}\ and\ \bibinfo {author} {\bibfnamefont {G.~M.}\ \bibnamefont
  {Zaslavsky}},\ }\href {\doibase 10.1016/0378-4371(78)90190-5} {\bibfield
  {journal} {\bibinfo  {journal} {Physica (Amsterdam)}\ }\textbf {\bibinfo
  {volume} {91A}},\ \bibinfo {pages} {450} (\bibinfo {year}
  {1978})}\BibitemShut {NoStop}%
\bibitem [{\citenamefont {Dvali}\ \emph {et~al.}(2013)\citenamefont {Dvali},
  \citenamefont {Flassig}, \citenamefont {Gomez}, \citenamefont {Pritzel},\
  and\ \citenamefont {Wintergerst}}]{DvaFlaGomPriPRD2013}%
  \BibitemOpen
  \bibfield  {author} {\bibinfo {author} {\bibfnamefont {G.}~\bibnamefont
  {Dvali}}, \bibinfo {author} {\bibfnamefont {D.}~\bibnamefont {Flassig}},
  \bibinfo {author} {\bibfnamefont {C.}~\bibnamefont {Gomez}}, \bibinfo
  {author} {\bibfnamefont {A.}~\bibnamefont {Pritzel}}, \ and\ \bibinfo
  {author} {\bibfnamefont {N.}~\bibnamefont {Wintergerst}},\ }\href {\doibase
  10.1103/PhysRevD.88.124041} {\bibfield  {journal} {\bibinfo  {journal}
  {Phys.~Rev.~D}\ }\textbf {\bibinfo {volume} {88}},\ \bibinfo {pages} {124041}
  (\bibinfo {year} {2013})}\BibitemShut {NoStop}%
\bibitem [{\citenamefont {Tomsovic}\ and\ \citenamefont
  {Heller}(1991)}]{TomHelPRL1991}%
  \BibitemOpen
  \bibfield  {author} {\bibinfo {author} {\bibfnamefont {S.}~\bibnamefont
  {Tomsovic}}\ and\ \bibinfo {author} {\bibfnamefont {E.~J.}\ \bibnamefont
  {Heller}},\ }\href {\doibase 10.1103/PhysRevLett.67.664} {\bibfield
  {journal} {\bibinfo  {journal} {Phys.~Rev.~Lett.}\ }\textbf {\bibinfo
  {volume} {67}},\ \bibinfo {pages} {664} (\bibinfo {year} {1991})}\BibitemShut
  {NoStop}%
\bibitem [{\citenamefont {Aleiner}\ and\ \citenamefont
  {Larkin}(1996)}]{AleLarPRB1996}%
  \BibitemOpen
  \bibfield  {author} {\bibinfo {author} {\bibfnamefont {I.~L.}\ \bibnamefont
  {Aleiner}}\ and\ \bibinfo {author} {\bibfnamefont {A.~I.}\ \bibnamefont
  {Larkin}},\ }\href {\doibase 10.1103/PhysRevB.54.14423} {\bibfield  {journal}
  {\bibinfo  {journal} {Phys.~Rev.~B}\ }\textbf {\bibinfo {volume} {54}},\
  \bibinfo {pages} {14423} (\bibinfo {year} {1996})}\BibitemShut {NoStop}%
\bibitem [{\citenamefont {Agam}\ \emph {et~al.}(2000)\citenamefont {Agam},
  \citenamefont {Aleiner},\ and\ \citenamefont {Larkin}}]{AgaAleLarPRL2000}%
  \BibitemOpen
  \bibfield  {author} {\bibinfo {author} {\bibfnamefont {O.}~\bibnamefont
  {Agam}}, \bibinfo {author} {\bibfnamefont {I.}~\bibnamefont {Aleiner}}, \
  and\ \bibinfo {author} {\bibfnamefont {A.}~\bibnamefont {Larkin}},\ }\href
  {\doibase 10.1103/PhysRevLett.85.3153} {\bibfield  {journal} {\bibinfo
  {journal} {Phys.~Rev.~Lett.}\ }\textbf {\bibinfo {volume} {85}},\ \bibinfo
  {pages} {3153} (\bibinfo {year} {2000})}\BibitemShut {NoStop}%
\bibitem [{\citenamefont {Sieber}\ and\ \citenamefont
  {Richter}(2001)}]{SieRicPS2001}%
  \BibitemOpen
  \bibfield  {author} {\bibinfo {author} {\bibfnamefont {M.}~\bibnamefont
  {Sieber}}\ and\ \bibinfo {author} {\bibfnamefont {K.}~\bibnamefont
  {Richter}},\ }\href {\doibase 10.1238/Physica.Topical.090a00128} {\bibfield
  {journal} {\bibinfo  {journal} {Physica Scripta}\ }\textbf {\bibinfo {volume}
  {T90}},\ \bibinfo {pages} {128} (\bibinfo {year} {2001})}\BibitemShut
  {NoStop}%
\bibitem [{\citenamefont {M{\"{u}}ller}\ \emph {et~al.}(2005)\citenamefont
  {M{\"{u}}ller}, \citenamefont {Heusler}, \citenamefont {Braun}, \citenamefont
  {Haake},\ and\ \citenamefont {Altland}}]{MulHeuBraHaaPRE2005}%
  \BibitemOpen
  \bibfield  {author} {\bibinfo {author} {\bibfnamefont {S.}~\bibnamefont
  {M{\"{u}}ller}}, \bibinfo {author} {\bibfnamefont {S.}~\bibnamefont
  {Heusler}}, \bibinfo {author} {\bibfnamefont {P.}~\bibnamefont {Braun}},
  \bibinfo {author} {\bibfnamefont {F.}~\bibnamefont {Haake}}, \ and\ \bibinfo
  {author} {\bibfnamefont {A.}~\bibnamefont {Altland}},\ }\href {\doibase
  10.1103/PhysRevE.72.046207} {\bibfield  {journal} {\bibinfo  {journal}
  {Phys.~Rev.~E}\ }\textbf {\bibinfo {volume} {72}},\ \bibinfo {pages} {046207}
  (\bibinfo {year} {2005})}\BibitemShut {NoStop}%
\bibitem [{\citenamefont {Brouwer}\ and\ \citenamefont
  {Rahav}(2006{\natexlab{a}})}]{BroRahPRB2006}%
  \BibitemOpen
  \bibfield  {author} {\bibinfo {author} {\bibfnamefont {P.~W.}\ \bibnamefont
  {Brouwer}}\ and\ \bibinfo {author} {\bibfnamefont {S.}~\bibnamefont
  {Rahav}},\ }\href {\doibase 10.1103/PhysRevB.74.075322} {\bibfield  {journal}
  {\bibinfo  {journal} {Phys.~Rev.~B}\ }\textbf {\bibinfo {volume} {74}},\
  \bibinfo {pages} {075322} (\bibinfo {year} {2006}{\natexlab{a}})}\BibitemShut
  {NoStop}%
\bibitem [{\citenamefont {Jacquod}\ and\ \citenamefont
  {Whitney}(2006)}]{JacWhiPRB2006}%
  \BibitemOpen
  \bibfield  {author} {\bibinfo {author} {\bibfnamefont {P.}~\bibnamefont
  {Jacquod}}\ and\ \bibinfo {author} {\bibfnamefont {R.~S.}\ \bibnamefont
  {Whitney}},\ }\href {\doibase 10.1103/PhysRevB.73.195115} {\bibfield
  {journal} {\bibinfo  {journal} {Phys.~Rev.~B}\ }\textbf {\bibinfo {volume}
  {73}},\ \bibinfo {pages} {195115} (\bibinfo {year} {2006})}\BibitemShut
  {NoStop}%
\bibitem [{\citenamefont {Waltner}\ \emph {et~al.}(2008)\citenamefont
  {Waltner}, \citenamefont {Guti{\'{e}}rrez}, \citenamefont {Goussev},\ and\
  \citenamefont {Richter}}]{WalGutGouRicPRL2008}%
  \BibitemOpen
  \bibfield  {author} {\bibinfo {author} {\bibfnamefont {D.}~\bibnamefont
  {Waltner}}, \bibinfo {author} {\bibfnamefont {M.}~\bibnamefont
  {Guti{\'{e}}rrez}}, \bibinfo {author} {\bibfnamefont {A.}~\bibnamefont
  {Goussev}}, \ and\ \bibinfo {author} {\bibfnamefont {K.}~\bibnamefont
  {Richter}},\ }\href {\doibase 10.1103/PhysRevLett.101.174101} {\bibfield
  {journal} {\bibinfo  {journal} {Phys.~Rev.~Lett.}\ }\textbf {\bibinfo
  {volume} {101}},\ \bibinfo {pages} {174101} (\bibinfo {year}
  {2008})}\BibitemShut {NoStop}%
\bibitem [{\citenamefont {Engl}\ \emph
  {et~al.}(2014{\natexlab{a}})\citenamefont {Engl}, \citenamefont {Dujardin},
  \citenamefont {Arg\"uelles}, \citenamefont {Schlagheck}, \citenamefont
  {Richter},\ and\ \citenamefont {Urbina}}]{EngDujArgSchlaRicUrb13PRL}%
  \BibitemOpen
  \bibfield  {author} {\bibinfo {author} {\bibfnamefont {T.}~\bibnamefont
  {Engl}}, \bibinfo {author} {\bibfnamefont {J.}~\bibnamefont {Dujardin}},
  \bibinfo {author} {\bibfnamefont {A.}~\bibnamefont {Arg\"uelles}}, \bibinfo
  {author} {\bibfnamefont {P.}~\bibnamefont {Schlagheck}}, \bibinfo {author}
  {\bibfnamefont {K.}~\bibnamefont {Richter}}, \ and\ \bibinfo {author}
  {\bibfnamefont {J.~D.}\ \bibnamefont {Urbina}},\ }\href {\doibase
  10.1103/PhysRevLett.112.140403} {\bibfield  {journal} {\bibinfo  {journal}
  {Phys.~Rev.~Lett.}\ }\textbf {\bibinfo {volume} {112}},\ \bibinfo {pages}
  {140403} (\bibinfo {year} {2014}{\natexlab{a}})}\BibitemShut {NoStop}%
\bibitem [{\citenamefont {Urbina}\ \emph {et~al.}(2016)\citenamefont {Urbina},
  \citenamefont {Kuipers}, \citenamefont {Matsumoto}, \citenamefont {Hummel},\
  and\ \citenamefont {Richter}}]{UrbKuiHumMatRicPRL2016}%
  \BibitemOpen
  \bibfield  {author} {\bibinfo {author} {\bibfnamefont {J.~D.}\ \bibnamefont
  {Urbina}}, \bibinfo {author} {\bibfnamefont {J.}~\bibnamefont {Kuipers}},
  \bibinfo {author} {\bibfnamefont {S.}~\bibnamefont {Matsumoto}}, \bibinfo
  {author} {\bibfnamefont {Q.}~\bibnamefont {Hummel}}, \ and\ \bibinfo {author}
  {\bibfnamefont {K.}~\bibnamefont {Richter}},\ }\href {\doibase
  10.1103/PhysRevLett.116.100401} {\bibfield  {journal} {\bibinfo  {journal}
  {Phys.~Rev.~Lett.}\ }\textbf {\bibinfo {volume} {116}},\ \bibinfo {pages}
  {100401} (\bibinfo {year} {2016})}\BibitemShut {NoStop}%
\bibitem [{\citenamefont {Engl}\ \emph {et~al.}(2015)\citenamefont {Engl},
  \citenamefont {Urbina},\ and\ \citenamefont {Richter}}]{EngUrbRicPRE2015}%
  \BibitemOpen
  \bibfield  {author} {\bibinfo {author} {\bibfnamefont {T.}~\bibnamefont
  {Engl}}, \bibinfo {author} {\bibfnamefont {J.~D.}\ \bibnamefont {Urbina}}, \
  and\ \bibinfo {author} {\bibfnamefont {K.}~\bibnamefont {Richter}},\ }\href
  {\doibase 10.1103/PhysRevE.92.062907} {\bibfield  {journal} {\bibinfo
  {journal} {Phys.~Rev.~E}\ }\textbf {\bibinfo {volume} {92}},\ \bibinfo
  {pages} {062907} (\bibinfo {year} {2015})}\BibitemShut {NoStop}%
\bibitem [{\citenamefont {Dubertrand}\ and\ \citenamefont
  {M{\"{u}}ller}(2016)}]{DubMulNJP2016}%
  \BibitemOpen
  \bibfield  {author} {\bibinfo {author} {\bibfnamefont {R.}~\bibnamefont
  {Dubertrand}}\ and\ \bibinfo {author} {\bibfnamefont {S.}~\bibnamefont
  {M{\"{u}}ller}},\ }\href {\doibase 10.1088/1367-2630/18/3/033009} {\bibfield
  {journal} {\bibinfo  {journal} {New J.~Phys.}\ }\textbf {\bibinfo {volume}
  {18}},\ \bibinfo {pages} {033009} (\bibinfo {year} {2016})}\BibitemShut
  {NoStop}%
\bibitem [{\citenamefont {Akila}\ \emph {et~al.}(2017)\citenamefont {Akila},
  \citenamefont {Waltner}, \citenamefont {Gutkin}, \citenamefont {Braun},\ and\
  \citenamefont {Guhr}}]{AkiWalGutBraGuhPRL2017}%
  \BibitemOpen
  \bibfield  {author} {\bibinfo {author} {\bibfnamefont {M.}~\bibnamefont
  {Akila}}, \bibinfo {author} {\bibfnamefont {D.}~\bibnamefont {Waltner}},
  \bibinfo {author} {\bibfnamefont {B.}~\bibnamefont {Gutkin}}, \bibinfo
  {author} {\bibfnamefont {P.}~\bibnamefont {Braun}}, \ and\ \bibinfo {author}
  {\bibfnamefont {T.}~\bibnamefont {Guhr}},\ }\href {\doibase
  10.1103/PhysRevLett.118.164101} {\bibfield  {journal} {\bibinfo  {journal}
  {Phys.~Rev.~Lett.}\ }\textbf {\bibinfo {volume} {118}},\ \bibinfo {pages}
  {164101} (\bibinfo {year} {2017})}\BibitemShut {NoStop}%
\bibitem [{\citenamefont {Tomsovic}\ \emph {et~al.}(2018)\citenamefont
  {Tomsovic}, \citenamefont {Schlagheck}, \citenamefont {Ullmo}, \citenamefont
  {Urbina},\ and\ \citenamefont {Richter}}]{TomSchlaUllUrbRic2018PRA}%
  \BibitemOpen
  \bibfield  {author} {\bibinfo {author} {\bibfnamefont {S.}~\bibnamefont
  {Tomsovic}}, \bibinfo {author} {\bibfnamefont {P.}~\bibnamefont
  {Schlagheck}}, \bibinfo {author} {\bibfnamefont {D.}~\bibnamefont {Ullmo}},
  \bibinfo {author} {\bibfnamefont {J.~D.}\ \bibnamefont {Urbina}}, \ and\
  \bibinfo {author} {\bibfnamefont {K.}~\bibnamefont {Richter}},\ }\href
  {\doibase 10.1103/PhysRevA.97.061606} {\bibfield  {journal} {\bibinfo
  {journal} {Phys.~Rev.~A}\ }\textbf {\bibinfo {volume} {97}},\ \bibinfo
  {pages} {061606} (\bibinfo {year} {2018})}\BibitemShut {NoStop}%
\bibitem [{\citenamefont {Agarwal}(2013)}]{Agarwal2013}%
  \BibitemOpen
  \bibfield  {author} {\bibinfo {author} {\bibfnamefont {G.~S.}\ \bibnamefont
  {Agarwal}},\ }\href
  {http://www.cambridge.org/de/academic/subjects/physics/optics-optoelectronics-and-photonics/quantum-optics-1}
  {\emph {\bibinfo {title} {{Quantum Optics}}}}\ (\bibinfo  {publisher}
  {Cambridge University Press},\ \bibinfo {address} {Cambridge, England},\
  \bibinfo {year} {2013})\BibitemShut {NoStop}%
\bibitem [{Sup()}]{SuppMat}%
  \BibitemOpen
  \href@noop {} {}\bibinfo {note} {{See the subsequent Supplemental Material
  for the detailed technical evaluation of the diagrams (a) to (d) in Fig.~1
  and the technical discussion of the generalization to other operators. It
  cites Refs.~\cite{Gas1998Book, MulHeuBraHaaPRE2005, WalGutGouRicPRL2008,
  MulNJP2007, TurRicJPA2003, SpeJPA2003,
  Waltner2012,BroRahPRB2006b,Schleich2001,BroRahPRB2006}}}\BibitemShut
  {NoStop}%
\bibitem [{\citenamefont {Adagideli}(2003)}]{AdaPRB2003}%
  \BibitemOpen
  \bibfield  {author} {\bibinfo {author} {\bibfnamefont {I.}~\bibnamefont
  {Adagideli}},\ }\href {\doibase 10.1103/PhysRevB.68.233308} {\bibfield
  {journal} {\bibinfo  {journal} {Phys.~Rev.~B}\ }\textbf {\bibinfo {volume}
  {68}},\ \bibinfo {pages} {233308} (\bibinfo {year} {2003})}\BibitemShut
  {NoStop}%
\bibitem [{\citenamefont {Brouwer}(2007)}]{BroPRB2007}%
  \BibitemOpen
  \bibfield  {author} {\bibinfo {author} {\bibfnamefont {P.~W.}\ \bibnamefont
  {Brouwer}},\ }\href {\doibase 10.1103/PhysRevB.76.165313} {\bibfield
  {journal} {\bibinfo  {journal} {Phys.~Rev.~B}\ }\textbf {\bibinfo {volume}
  {76}},\ \bibinfo {pages} {165313} (\bibinfo {year} {2007})}\BibitemShut
  {NoStop}%
\bibitem [{\citenamefont {Kuipers}\ \emph {et~al.}(2010)\citenamefont
  {Kuipers}, \citenamefont {Waltner}, \citenamefont {Petitjean}, \citenamefont
  {Berkolaiko},\ and\ \citenamefont {Richter}}]{KuiWalPetBerRicPRL2010}%
  \BibitemOpen
  \bibfield  {author} {\bibinfo {author} {\bibfnamefont {J.}~\bibnamefont
  {Kuipers}}, \bibinfo {author} {\bibfnamefont {D.}~\bibnamefont {Waltner}},
  \bibinfo {author} {\bibfnamefont {C.}~\bibnamefont {Petitjean}}, \bibinfo
  {author} {\bibfnamefont {G.}~\bibnamefont {Berkolaiko}}, \ and\ \bibinfo
  {author} {\bibfnamefont {K.}~\bibnamefont {Richter}},\ }\href {\doibase
  10.1103/PhysRevLett.104.027001} {\bibfield  {journal} {\bibinfo  {journal}
  {Phys.~Rev.~Lett.}\ }\textbf {\bibinfo {volume} {104}},\ \bibinfo {pages}
  {027001} (\bibinfo {year} {2010})}\BibitemShut {NoStop}%
\bibitem [{Note2()}]{Note2}%
  \BibitemOpen
  \bibinfo {note} {Specific diagrams similar to class (d) in Fig.~\ref
  {fig:OTOC_diagram_classes} have been considered in the context of shot noise
  \cite {Lassl2003, BraHeuMulHaaJPA2006, BroPRB2007} and quantum chaotic SP
  \cite {KuiWalPetBerRicPRL2010} and MB \cite {UrbKuiHumMatRicPRL2016}
  scattering.}\BibitemShut {Stop}%
\bibitem [{\citenamefont {Ozorio~de Almeida}(1990)}]{Ozo1990}%
  \BibitemOpen
  \bibfield  {author} {\bibinfo {author} {\bibfnamefont {A.~M.}\ \bibnamefont
  {Ozorio~de Almeida}},\ }\href
  {http://www.cambridge.org/academic/subjects/physics/nonlinear-science-and-fluid-dynamics/hamiltonian-systems-chaos-and-quantization}
  {\emph {\bibinfo {title} {{Hamiltonian Systems: Chaos and Quantization}}}}\
  (\bibinfo  {publisher} {Cambridge University Press},\ \bibinfo {address}
  {Cambridge, England},\ \bibinfo {year} {1990})\BibitemShut {NoStop}%
\bibitem [{\citenamefont {Gaspard}(1998)}]{Gas1998Book}%
  \BibitemOpen
  \bibfield  {author} {\bibinfo {author} {\bibfnamefont {P.}~\bibnamefont
  {Gaspard}},\ }\href {\doibase 10.1017/CBO9780511628856} {\emph {\bibinfo
  {title} {Chaos, Scattering and Statistical Mechanics}}}\ (\bibinfo
  {publisher} {Cambridge University Press},\ \bibinfo {address} {Cambridge,
  England},\ \bibinfo {year} {1998})\BibitemShut {NoStop}%
\bibitem [{\citenamefont {Han}\ and\ \citenamefont {Wu}(2016)}]{HanWuPRA2016}%
  \BibitemOpen
  \bibfield  {author} {\bibinfo {author} {\bibfnamefont {X.}~\bibnamefont
  {Han}}\ and\ \bibinfo {author} {\bibfnamefont {B.}~\bibnamefont {Wu}},\
  }\href {\doibase 10.1103/PhysRevA.93.023621} {\bibfield  {journal} {\bibinfo
  {journal} {Phys.~Rev.~A}\ }\textbf {\bibinfo {volume} {93}},\ \bibinfo
  {pages} {023621} (\bibinfo {year} {2016})}\BibitemShut {NoStop}%
\bibitem [{Note3()}]{Note3}%
  \BibitemOpen
  \bibinfo {note} {It may be viewed as the underlying dynamical mechanism,
  supporting (in the large-$N$ limit) models for OTOCs based on coupled
  binaries~\cite {RakPolKey2018PRX}.}\BibitemShut {Stop}%
\bibitem [{Note4()}]{Note4}%
  \BibitemOpen
  \bibinfo {note} {Ehrenfest time effects in Lorentz gases were studied,
  {\protect \it e.g.}, in Refs.~\cite {AleLarPRB1996, YevLutWeiRicPRL2000,
  BroPRB2007}.}\BibitemShut {Stop}%
\bibitem [{\citenamefont {Borgonovi}\ \emph {et~al.}(2018)\citenamefont
  {Borgonovi}, \citenamefont {Izrailev},\ and\ \citenamefont
  {Santos}}]{BorIzrSanarxiv2018}%
  \BibitemOpen
  \bibfield  {author} {\bibinfo {author} {\bibfnamefont {F.}~\bibnamefont
  {Borgonovi}}, \bibinfo {author} {\bibfnamefont {F.~M.}\ \bibnamefont
  {Izrailev}}, \ and\ \bibinfo {author} {\bibfnamefont {L.~F.}\ \bibnamefont
  {Santos}},\ }\href@noop {} {} (\bibinfo {year} {2018}),\ \Eprint
  {http://arxiv.org/abs/1802.08265} {arXiv:1802.08265} \BibitemShut {NoStop}%
\bibitem [{\citenamefont {Tomsovic}(2018)}]{Tom2018PRE}%
  \BibitemOpen
  \bibfield  {author} {\bibinfo {author} {\bibfnamefont {S.}~\bibnamefont
  {Tomsovic}},\ }\href {\doibase 10.1103/PhysRevE.98.023301} {\bibfield
  {journal} {\bibinfo  {journal} {Phys.~Rev.~E}\ }\textbf {\bibinfo {volume}
  {98}},\ \bibinfo {pages} {023301} (\bibinfo {year} {2018})}\BibitemShut
  {NoStop}%
\bibitem [{Note5()}]{Note5}%
  \BibitemOpen
  \bibinfo {note} {To this end, the semiclassical (large-$N$) approximation for
  the microscopic path integral propagator of discrete fermionic quantum
  fields~\cite {EngPloUrbRic2014TCA} can be employed. Based on this fermionic
  propagator, a semiclassical calculation of a MB spin echo gave perfect
  agreement with numerical quantum calculations, see Ref.~\cite
  {EngUrbRicSchla2014PRA}.}\BibitemShut {Stop}%
\bibitem [{\citenamefont {M{\"{u}}ller}\ \emph {et~al.}(2007)\citenamefont
  {M{\"{u}}ller}, \citenamefont {Heusler}, \citenamefont {Braun},\ and\
  \citenamefont {Haake}}]{MulNJP2007}%
  \BibitemOpen
  \bibfield  {author} {\bibinfo {author} {\bibfnamefont {S.}~\bibnamefont
  {M{\"{u}}ller}}, \bibinfo {author} {\bibfnamefont {S.}~\bibnamefont
  {Heusler}}, \bibinfo {author} {\bibfnamefont {P.}~\bibnamefont {Braun}}, \
  and\ \bibinfo {author} {\bibfnamefont {F.}~\bibnamefont {Haake}},\ }\href
  {\doibase 10.1088/1367-2630/9/1/012} {\bibfield  {journal} {\bibinfo
  {journal} {New J.~Phys.}\ }\textbf {\bibinfo {volume} {9}},\ \bibinfo {pages}
  {12} (\bibinfo {year} {2007})}\BibitemShut {NoStop}%
\bibitem [{\citenamefont {Turek}\ and\ \citenamefont
  {Richter}(2003)}]{TurRicJPA2003}%
  \BibitemOpen
  \bibfield  {author} {\bibinfo {author} {\bibfnamefont {M.}~\bibnamefont
  {Turek}}\ and\ \bibinfo {author} {\bibfnamefont {K.}~\bibnamefont
  {Richter}},\ }\href {\doibase 10.1088/0305-4470/36/30/101} {\bibfield
  {journal} {\bibinfo  {journal} {J.~Phys.~A}\ }\textbf {\bibinfo {volume}
  {36}},\ \bibinfo {pages} {L455} (\bibinfo {year} {2003})}\BibitemShut
  {NoStop}%
\bibitem [{\citenamefont {Spehner}(2003)}]{SpeJPA2003}%
  \BibitemOpen
  \bibfield  {author} {\bibinfo {author} {\bibfnamefont {D.}~\bibnamefont
  {Spehner}},\ }\href {\doibase 10.1088/0305-4470/36/26/304} {\bibfield
  {journal} {\bibinfo  {journal} {J.~Phys.~A}\ }\textbf {\bibinfo {volume}
  {36}},\ \bibinfo {pages} {7269} (\bibinfo {year} {2003})}\BibitemShut
  {NoStop}%
\bibitem [{\citenamefont {Waltner}(2012)}]{Waltner2012}%
  \BibitemOpen
  \bibfield  {author} {\bibinfo {author} {\bibfnamefont {D.}~\bibnamefont
  {Waltner}},\ }\href {https://www.springer.com/us/book/9783642245275} {\emph
  {\bibinfo {title} {{Semiclassical Approach to Mesoscopic Systems: Classical
  Trajectory Correlations and Wave Interference}}}}\ (\bibinfo  {publisher}
  {Springer, Berlin, Heidelberg},\ \bibinfo {year} {2012})\BibitemShut
  {NoStop}%
\bibitem [{\citenamefont {Brouwer}\ and\ \citenamefont
  {Rahav}(2006{\natexlab{b}})}]{BroRahPRB2006b}%
  \BibitemOpen
  \bibfield  {author} {\bibinfo {author} {\bibfnamefont {P.~W.}\ \bibnamefont
  {Brouwer}}\ and\ \bibinfo {author} {\bibfnamefont {S.}~\bibnamefont
  {Rahav}},\ }\href {\doibase 10.1103/PhysRevB.74.085313} {\bibfield  {journal}
  {\bibinfo  {journal} {Phys.~Rev.~B}\ }\textbf {\bibinfo {volume} {74}},\
  \bibinfo {pages} {085313} (\bibinfo {year} {2006}{\natexlab{b}})}\BibitemShut
  {NoStop}%
\bibitem [{\citenamefont {Schleich}(2001)}]{Schleich2001}%
  \BibitemOpen
  \bibfield  {author} {\bibinfo {author} {\bibfnamefont {W.}~\bibnamefont
  {Schleich}},\ }\href
  {https://onlinelibrary.wiley.com/doi/book/10.1002/3527602976} {\emph
  {\bibinfo {title} {{Quantum Optics in Phase Space}}}}\ (\bibinfo  {publisher}
  {Wiley-VCH},\ \bibinfo {address} {Weinheim, Germany},\ \bibinfo {year}
  {2001})\BibitemShut {NoStop}%
\bibitem [{\citenamefont {Lassl}(2003)}]{Lassl2003}%
  \BibitemOpen
  \bibfield  {author} {\bibinfo {author} {\bibfnamefont {A.}~\bibnamefont
  {Lassl}},\ }\emph {\bibinfo {title} {{Semiklassik jenseits der
  Diagonaln{\"{a}}herung: Anwendung auf ballisitische Systeme}}},\ \href
  {http://epub.uni-regensburg.de/1521/} {\bibinfo {type} {{Diploma thesis}}},\
  \bibinfo  {school} {Universit{\"{a}}t Regensburg} (\bibinfo {year}
  {2003})\BibitemShut {NoStop}%
\bibitem [{\citenamefont {Braun}\ \emph {et~al.}(2006)\citenamefont {Braun},
  \citenamefont {Heusler}, \citenamefont {M{\"{u}}ller},\ and\ \citenamefont
  {Haake}}]{BraHeuMulHaaJPA2006}%
  \BibitemOpen
  \bibfield  {author} {\bibinfo {author} {\bibfnamefont {P.}~\bibnamefont
  {Braun}}, \bibinfo {author} {\bibfnamefont {S.}~\bibnamefont {Heusler}},
  \bibinfo {author} {\bibfnamefont {S.}~\bibnamefont {M{\"{u}}ller}}, \ and\
  \bibinfo {author} {\bibfnamefont {F.}~\bibnamefont {Haake}},\ }\href
  {\doibase 10.1088/0305-4470/39/11/L01} {\bibfield  {journal} {\bibinfo
  {journal} {J.~Phys.~A}\ }\textbf {\bibinfo {volume} {39}},\ \bibinfo {pages}
  {L159} (\bibinfo {year} {2006})}\BibitemShut {NoStop}%
\bibitem [{\citenamefont {Rakovszky}\ \emph {et~al.}(2018)\citenamefont
  {Rakovszky}, \citenamefont {Pollmann},\ and\ \citenamefont {von
  Keyserlingk}}]{RakPolKey2018PRX}%
  \BibitemOpen
  \bibfield  {author} {\bibinfo {author} {\bibfnamefont {T.}~\bibnamefont
  {Rakovszky}}, \bibinfo {author} {\bibfnamefont {F.}~\bibnamefont {Pollmann}},
  \ and\ \bibinfo {author} {\bibfnamefont {C.~W.}\ \bibnamefont {von
  Keyserlingk}},\ }\href {\doibase 10.1103/PhysRevX.8.031058} {\bibfield
  {journal} {\bibinfo  {journal} {Phys.~Rev.~X}\ }\textbf {\bibinfo {volume}
  {8}},\ \bibinfo {pages} {031058} (\bibinfo {year} {2018})}\BibitemShut
  {NoStop}%
\bibitem [{\citenamefont {Yevtushenko}\ \emph {et~al.}(2000)\citenamefont
  {Yevtushenko}, \citenamefont {L{\"{u}}tjering}, \citenamefont {Weiss},\ and\
  \citenamefont {Richter}}]{YevLutWeiRicPRL2000}%
  \BibitemOpen
  \bibfield  {author} {\bibinfo {author} {\bibfnamefont {O.}~\bibnamefont
  {Yevtushenko}}, \bibinfo {author} {\bibfnamefont {G.}~\bibnamefont
  {L{\"{u}}tjering}}, \bibinfo {author} {\bibfnamefont {D.}~\bibnamefont
  {Weiss}}, \ and\ \bibinfo {author} {\bibfnamefont {K.}~\bibnamefont
  {Richter}},\ }\href {\doibase 10.1103/PhysRevLett.84.542} {\bibfield
  {journal} {\bibinfo  {journal} {Phys.~Rev.~Lett.}\ }\textbf {\bibinfo
  {volume} {84}},\ \bibinfo {pages} {542} (\bibinfo {year} {2000})}\BibitemShut
  {NoStop}%
\bibitem [{\citenamefont {Engl}\ \emph
  {et~al.}(2014{\natexlab{b}})\citenamefont {Engl}, \citenamefont
  {Pl{\"{o}}{\ss}l}, \citenamefont {Urbina},\ and\ \citenamefont
  {Richter}}]{EngPloUrbRic2014TCA}%
  \BibitemOpen
  \bibfield  {author} {\bibinfo {author} {\bibfnamefont {T.}~\bibnamefont
  {Engl}}, \bibinfo {author} {\bibfnamefont {P.}~\bibnamefont
  {Pl{\"{o}}{\ss}l}}, \bibinfo {author} {\bibfnamefont {J.~D.}\ \bibnamefont
  {Urbina}}, \ and\ \bibinfo {author} {\bibfnamefont {K.}~\bibnamefont
  {Richter}},\ }\href {\doibase 10.1007/s00214-014-1563-9} {\bibfield
  {journal} {\bibinfo  {journal} {Theor.~Chem.~Acc.}\ }\textbf {\bibinfo
  {volume} {133}},\ \bibinfo {pages} {1563} (\bibinfo {year}
  {2014}{\natexlab{b}})}\BibitemShut {NoStop}%
\bibitem [{\citenamefont {Engl}\ \emph {et~al.}(2018)\citenamefont {Engl},
  \citenamefont {Urbina}, \citenamefont {Richter},\ and\ \citenamefont
  {Schlagheck}}]{EngUrbRicSchla2014PRA}%
  \BibitemOpen
  \bibfield  {author} {\bibinfo {author} {\bibfnamefont {T.}~\bibnamefont
  {Engl}}, \bibinfo {author} {\bibfnamefont {J.~D.}\ \bibnamefont {Urbina}},
  \bibinfo {author} {\bibfnamefont {K.}~\bibnamefont {Richter}}, \ and\
  \bibinfo {author} {\bibfnamefont {P.}~\bibnamefont {Schlagheck}},\ }\href
  {\doibase 10.1103/PhysRevA.98.013630} {\bibfield  {journal} {\bibinfo
  {journal} {Phys.~Rev.~A}\ }\textbf {\bibinfo {volume} {98}},\ \bibinfo
  {pages} {013630} (\bibinfo {year} {2018})}\BibitemShut {NoStop}%
\end{thebibliography}%

\cleardoublepage

\onecolumngrid

\begin{center}
  \textbf{\Large{Supplemental material to the paper \\[1ex]
      ``Many-Body Quantum Interference and the Saturation\\ of
      Out-of-Time-Order Correlators''}}
\end{center}
\begin{center}
  Josef Rammensee$,^1$ Juan Diego Urbina$,^1$ and Klaus Richter$^1$\\
  \emph{\small $^1$Institut f\"ur Theoretische Physik, Universit\"at
    Regensburg, D-93040 Regensburg, Germany }
\end{center}
\twocolumngrid

Here we provide detailed calculations of the contributions of the
diagram classes (a) to (d) (in Fig.~\ref{fig:OTOC_diagram_classes} of
the main text) to the out-of-time-order correlator (OTOC).

\section{Phase space structure of the classical limit of the
  generalized Bose-Hubbard system}
The classical limit of the Bose-Hubbard model described by Hamiltonian
(\ref{eq:H}) of the main text is found to be
$\mathcal{H}^{\mathrm{cl}}$ given in Eq.~(\ref{eq:H_cl}).
Generically, this Hamiltonian has at least two constants of motion
(CoM), the energy, represented by the value of
$\mathcal{H}^{\mathrm{cl}}$, and the conservation of particle density,
represented by
\begin{equation}
  \mathcal{N}(\vec{q},\vec{p})
  = \sum_{i=1}^n|\Phi_i|^2 = \frac{1}{2}\sum_{i=1}^n\left(q_i^2+p_i^2\right)
  \,.
  \label{eq:particle_density}
\end{equation}
We assume that the system does not have any other CoMs and displays
fully chaotic motion on the $(2n-2)$-dimensional submanifold defined
by the CoMs.
Locally at any phase-space (PS) point $\vec{x}=(\vec{q},\vec{p})$, the
coordinate system of the tangent space can be defined in such a way
that for each CoM one can associate 2 axes (parallel and perpendicular
to the flow defined by them), and the remaining $2(n-2)$ axis point
along the stable or unstable directions responsible for the fully
hyperbolic dynamics, see \emph{e.g.}~\cite{Gas1998Book}.
In the following, we denote the latter directions with
$\veces{l}{\vec{x}}$, $\veceu{l}{\vec{x}}$, $l\!=\!1,\ldots
n-2$.
Physically, if the difference of the initial conditions of two
trajectories lies in the direction $\veces{l}{\vec{x}}$
$(\veceu{l}{\vec{x}})$, the hyperbolic dynamics of the chaotic system
will exponentially increase (decrease) this difference, with a rate
given by the associated classical Lyapunov exponent $\lambda_l$.
For simplicity, we will assume that the chaotic dynamics is uniformly
hyperbolic, \emph{i.e.}~all stable and unstable directions share the
same Lyapunov exponent $\lambda$. A discussion of the generic
hyperbolic case would only lead to a significant increase of the
complexity of the calculations, while the result remains the same in
the two limits $t\ll\tauE$ and $t\gg\tauE$.

\section{Geometry of encounters in phase space}
For our calculations it is necessary to understand how to quantify
constellations of two trajectories which encounter each other in a PS
region, as displayed in Fig.~\ref{fig:OTOC_diagram_classes}.
For a detailed analysis of the single-particle case, there is a broad
literature available \cite{MulNJP2007, MulHeuBraHaaPRE2005,
  WalGutGouRicPRL2008, TurRicJPA2003, SpeJPA2003}.
For the ease of the reader, and also to be able to explain some
OTOC-related special aspects in the next section, we summarize the key
steps here.

The main idea is that during an encounter of two trajectories in PS
the dynamics of their relative motion is well described by linearizing
Hamilton's equations of motion around one of the trajectories.
In this linearized regime, the relative difference of the trajectories
in PS can be expressed in the local coordinate system spanned by the
directions towards the stable and unstable manifolds, as well as the
manifolds given by the CoMs, see the previous section.
However, we have to demand that both trajectories have (within a
window of $\mathcal{O}(\heff)$) the same values for their CoMs, since
later we will construct partner trajectories partially following both
trajectories.
Thus, the relative difference vector is expressed solely in terms of
the $2(n\!-\!2)$ stable and unstable directions.

To quantitatively describe two trajectories $\alpha$, $\beta$
encountering each other in PS, we first choose one of the trajectories
as a reference trajectory, say $\beta$, and then take a time $t'$ at
which we assume $\beta$ to be close to $\alpha$.
At the PS point of $\beta$ at $t'$, denoted by $\vec{x}_{\beta}(t')$,
we place the origin of a $2(n\!-\!2)$ dimensional coordinate system, which
is spanned by the local stable and unstable directions
$\veces[\beta]{l}{t'} \!\equiv\! \veces{l}{\vec{x}_{\beta}(t')}$,
$\veceu[\beta]{l}{t'} \!\equiv\! \veceu{l}{\vec{x}_{\beta}(t')}$.
In this frame, an encountering trajectory $\alpha$, which takes the
same values of the CoMs as $\beta$, is uniquely defined by vectors
$\vec{s}$, $\vec{u}$, as
\vspace*{-3ex}
\begin{equation}
  \vec{x}_{\alpha}(t')=\vec{x}_\beta(t') +\sum_{l=1}^{n-2}
  \left[s_l\veces[\beta]{l}{t'}+u_l\veceu[\beta]{l}{t'}
  \right]
  \vspace{-2ex}
  \label{eq:PS_point_alpha}
\end{equation}
uniquely defines the trajectory's PS point at time $t'$.
In the linearizable regime of the relative Hamiltonian dynamics,
\emph{i.e.}~as long as the components of the vectors $\vec{s}$,
$\vec{u}$ do not reach a given critical (classical) value $\pm c$,
this single PS point is well defined and, for a time-independent
Hamilton function, is sufficient to define the trajectory $\alpha$.
In the main text, this cutoff $c$ has been set to $1$ for the ease of
readability.
The only assumption is that $c^2$ is a typical classical
action scale, \emph{i.e.}~large compared to $\heff$.
Its exact value is not of importance as diagrams with action
differences much larger than $\heff$ do not contribute to the results
of semiclassical calculations, and reliable quantitative results do
not depend on it.

Note that in Eq.~(\ref{eq:PS_point_alpha}) the temporal
parametrization of $\alpha$ is such, that $\alpha$ enters the
encounter region simultaneously with $\beta$.
As seen from Eq.~(\ref{eq:OTOC_sc_integral_representation}), $\alpha$
and $\beta$ need the same time to get from the initial to the final
point.
A mismatch in times of the encounter event of the trajectories
$\alpha$ and $\beta$ would lead to partner trajectories $\alpha'$,
$\beta'$ with times different to $t$.
But those are not available in the sums over trajectories $\alpha'$,
$\beta'$ in Eq.~(\ref{eq:OTOC_sc_integral_representation}).

Subject to the hyperbolic dynamics, the vectors $\vec{s}$ and
$\vec{u}$ in the co-traveling coordinate system will change when
varying $t'$.
For instance for $t'$, $t''$ inside the encounter, the PS points given
by $\left( t', \vec{s}, \vec{u} \right)$ and
$( t'', \vec{s}\exp[-\lambda(t''\!-\!t')], \vec{u} \exp[\lambda(t''\!-\!t')])$
are describing the very same trajectory $\alpha$.
To avoid overcounting, it is necessary to later divide the
contributions by the time the trajectories spend inside the encounter
region.
The limits of the encounter regions are reached, when the first
components of $\vec{s}$ and $\vec{u}$ have grown to a classical scale
$\pm c$ at which the linearization breaks down.
This introduces two time scales,
\begin{align}
  \begin{aligned}
    \ts(\vec{s})
    &=\frac{1}{\lambda}
    \log\left(\frac{c}{\max_{i=1,\ldots,n-2}(|s_i|)}\right)\,,\\
    \tu(\vec{u})
    &=\frac{1}{\lambda}
    \log\left(\frac{c}{\max_{i=1,\ldots,n-2}(|u_i|)}\right)\,,
  \end{aligned}
      \label{eq:stable_unstable_times}
\end{align}
and the time for a fully developed encounter, as seen in
Fig.~\ref{fig:OTOC_diagram_classes} (d), is defined as the sum
\begin{equation}
  \tenc\left(\vec{s},\vec{u}\right)=  \ts(\vec{s}) +   \tu(\vec{u})\,.
  \label{eq:encounter_time}
\end{equation}
Note that if the trajectories start and/or end inside the encounter
region, as in Fig.~\ref{fig:OTOC_diagram_classes} (a) to (c), the
encounter time has to be reduced accordingly.

Suitable partner trajectories following the original trajectories
outside the encounter region, while interchanging partners inside it, are
found by the PS points
\begin{align}
  \begin{aligned}
  \vec{x}_{\beta'}(t')
  & = \vec{x}_\beta(t') + \sum_{l=1}^{n-2} s_l\veces[\beta]{l}{t'}\,,\\
  \vec{x}_{\alpha'}(t')
  & = \vec{x}_\beta(t') + \sum_{l=1}^{n-2} u_l\veceu[\beta]{l}{t'}\,.
  \end{aligned}
  \label{eq:partner_definitions}
\end{align}
According to their definition, trajectory $\beta'$ exponentially
approaches $\beta$ for times larger than $t'$, as their difference is
solely along stable directions.
For times smaller than $t'$, we have to consider time-reversed
dynamics, and the stable and unstable manifolds interchange their
roles.
Thus, for times smaller than $t'$, $\beta'$ exponentially separates
from $\beta$, and exponentially approaches $\alpha$ in the same
fashion, since
\begin{equation}
  \vec{x}_{\alpha}(t')-\vec{x}_{\beta'}(t')
  =
  \sum_{l=1}^{n-2}u_l\veceu[\beta]{l}{t'}.
\end{equation}
The same reasoning can be applied to $\alpha'$.

To summarize, a constellation of trajectories with a single encounter
is described by choosing one of the trajectories as a reference
trajectory, a time $t'$ as time of the encounter, and vectors
$\vec{s}$, $\vec{u}$ to quantify the respective distances towards the
other trajectories.

Regarding encounter contributions to OTOCs, there is a further
subtlety to consider.
If the initial points of the trajectories are contained inside the
encounter region, we have to treat classical quantities related to
initial points of trajectories in a correlated way, and also use the
local coordinates $\vec{s}$, $\vec{u}$ to describe them.
In Fig.~\ref{fig:OTOC_diagram_classes} (a) and (b), the beginning of
the trajectories is inside the encounter region. This requires to
treat the difference of initial momenta in
Eq.~(\ref{eq:OTOC_sc_integral_representation}) through
\begin{align}
  \begin{aligned}
    \init{p}_{\alpha',i} - \init{p}_{\alpha,i}
    &= - \sum_{l=1}^{n-2} s_l\rme^{\lambda t'}
    \left[\veces[\beta]{l}{0}\right]_{p_i},\\
    \init{p}_{\beta,i} - \init{p}_{\beta',i}
    &= - \sum_{l=1}^{n-2} s_l\rme^{\lambda t'}
    \left[\veces[\beta]{l}{0}\right]_{p_i}.
  \end{aligned}
      \label{eq:initial_momentum_difference}
\end{align}
Similarly, if the final points enter the encounter region, as in
Fig.~\ref{fig:OTOC_diagram_classes} (a) and (c), we use
\begin{align}
    &\fin{q}_{\alpha,j}\fin{q}_{\beta,j}
     = \frac{1}{2}\left( {\fin{q}_{\alpha,j}}^2 + {\fin{q}_{\beta,j}}^2\right)
    -\frac{1}{2}\left(\fin{q}_{\alpha,j} - \fin{q}_{\beta,j}\right)^2
    \nonumber
    \\
    & \approx
    {\fin{q}_{\alpha,j}}^2 -\frac{1}{2}
    \left( \sum_{l=1}^{n-2} u_l\rme^{\lambda (t-t')}
      \left[\veceu[\beta]{l}{t}\right]_{q_j}\right)^2.
      \label{eq:final_position_difference}
\end{align}
Here $\left[.\right]_{p_i}$ and $\left[.\right]_{q_j}$ denote the
$i$-th component of the momentum sector, and the $j$-th component of
the coordinate sector of the PS vector $\veceu[\beta]{l}{t}$.
As we will later approximate the square of the final points in
Eq.~(\ref{eq:final_position_difference}) by its ergodic average, we
already approximate them here by ${\fin{q}_{\alpha,j}}^2$ to simplify
the expressions.

\section{Density and action difference of diagrams with encounters}
To obtain all possible contributions to
Eq.~(\ref{eq:OTOC_sc_integral_representation}) from trajectory
constellations with an encounter, we first introduce integrations over
the relative differences $\vec{s}$, $\vec{u}$ and time $t'$ at which
these differences are employed.
The four-fold sum over trajectories is then reduced to a two-fold sum,
as the partner trajectories $\alpha'$ and $\beta'$ are uniquely given
by the Eqs.~(\ref{eq:partner_definitions}).
Furthermore, we correlate the remaining sums over $\alpha$ and $\beta$
by introducing the density distribution
\begin{equation}
  \rho_{\alpha,\beta}\left(\vec{s},\vec{u},t'\right)
  =
  \frac{(2\pi\heff)^2}{\tenc\left(\vec{s},\vec{u}\right)}
  \delta^{2n}
  \left[
    \vec{x}_\alpha(t') -
      \tilde{\vec{x}}\left(  \vec{x}_{\beta}(t'),\vec{s},\vec{u}\right)
    \right],
    \label{eq:density_of_encounters}
\end{equation}
where
\begin{equation}
  \tilde{\vec{x}}\left(  \vec{x},\vec{s},\vec{u}\right)
  =
    \vec{x} +
    \sum_{l=1}^{n-2}\left[
    s_l\veces{l}{\vec{x}}
    +
    u_l\veceu{l}{\vec{x}}
    \right]\,.
\end{equation}
The normalization $(2\pi\heff)^2$ in
Eq.~(\ref{eq:density_of_encounters}) is independently determined by
performing subsequent calculations imposing unitarity for the object
$ 1\!=\!\braket{\Psi| \hat{U}^\dagger(t) \hat{U}(t) \hat{U}^\dagger(t)
  \hat{U}(t) |\Psi}$.
It reflects that the paired trajectories should all stay in the window
of a Planck cell near the submanifold defined by the reference
trajectory's values for the CoMs energy and particle density.

The action difference of this system of four trajectories is found to be
\cite{TurRicJPA2003, MulHeuBraHaaPRE2005}
\begin{equation}
  R_{\alpha}-  R_{\alpha'} +  R_{\beta}-  R_{\beta'}
  \approx \vec{s}\cdot\vec{u} +\vecinit{p}_{\alpha} (\vec{q}_1-\vec{q}_5),
\end{equation}
where the latter term related to the initial momentum and the relative
distance $\vec{y}\!=\!\vec{q}_1\!-\!\vec{q}_5$ is introduced as we
substitute the trajectories $\alpha$, $\alpha'$, starting at
$\vec{q}_1$ and $\vec{q}_5$, by nearby trajectories starting at
$\vec{q}\!=\!\frac{1}{2}(\vec{q}_1\!+\!\vec{q}_5)$.

\section{Contributions of encounter diagrams to the OTOC}
\subsection{Contributions of 4-leg-encounters}
We start with the contributions of the 4-leg-encounters displayed in
Fig.~\ref{fig:OTOC_diagram_classes} (d).
This term is given by
\begin{widetext}
  \begin{align}
    C^{(\mathrm{4le})}(t)
    &=
    \int \rmd^n q\!
    \int \rmd^n y\!
    \int \rmd^n q_2\!
    \int\rmd^n q_3\!
    \int\rmd^n q_4\,
    \Psi^{*}\left(\vec{q}+ \frac{\vec{y}}{2}\right)
    \Psi\left(\vec{q}- \frac{\vec{y}}{2}\right)
    \hspace{-2ex}
    \sum_{
      \substack{
        \alpha: \vec{q}_3 \overset{t}{\rightarrow}\vec{q}_2\\
        \beta : \vec{q} \rightarrow\vec{q}_4}
    }\hspace{-2ex}
    |A_{\alpha}|^2 | A_{\beta}|^2
    \left(\init{p}_{\beta,i}-\init{p}_{\alpha,i} \right)^2
    \fin{q}_{\alpha,j} \fin{q}_{\beta,j}
    \rme^{\frac{\rmi}{\heff}\vecinit{p}_\alpha\vec{y}}
    \nonumber\\
    &\qquad\times
      \int_{-c}^c\rmd^{n-2} s\int_{-c}^c  \rmd^{n-2} u
      \int_{\ts(\vec{s})}^{t-\tu(\vec{u})}\rmd t'
      \rme^{ \frac{\rmi}{\heff}\vec{s}\vec{u}}
      \Theta\left[t-\tenc(\vec{s},\vec{u})\right]
      \rho_{\alpha,\beta}\left(\vec{s},\vec{u},t'\right).
      \label{eq:4le_starting_formula}
  \end{align}
\end{widetext}
Most of the ingredients for this integral have been already discussed
in the previous two sections.
The special features of Fig.~\ref{fig:OTOC_diagram_classes} (d) are
represented by the boundaries of the integration over $t'$, which
require that the encounter region does neither contain the beginning
nor the end of the trajectories.
The Heaviside step function $\Theta$ finally ensures that
encounter regions longer than the available time $t$ are excluded.

In a first step we use the fact that the squared amplitudes
$|A_\alpha|^2$ can be interpreted as Jacobian for a variable
transformation from final coordinates to initial momenta along a
classical trajectory,
\begin{equation}
  |A_\alpha|^2
  = \frac{1}{(2\pi\heff)^n}
  \left|\frac{\partial \vecinit{p}_{\alpha}}{\partial \vecfin{q}}\right|\,.
  \label{eq:stability_amplitude_squared}
\end{equation}
Together with the sum over trajectories $\alpha$, we can transform the
integrations over $\vec{q}_2\!=\!\vecfin{q}_{\alpha}$ to an
integration over initial momenta $\vec{p}_3$.
Trajectory-related quantities labeled by $\alpha$ become then
functions of trajectories with initial conditions
$\vec{x}_3\!=\!(\vec{q}_3,\vec{p}_3)$, \emph{e.g.}
\begin{equation}
  \left(\vecinit{p}_\alpha, \vecfin{q}_{\alpha}\right)
  \to
  \left( \vec{p}_3, \vecfin{q}\left(\vec{q}_3,\vec{p}_3;t\right)\right)\,.
\end{equation}
In the same spirit we use the sum over $\beta$ with $|A_\beta|^2$ to
transform the integration over $\vec{q}_4$ to $\vec{p}$, and
$\beta$-labeled quantities become functions of
$\vec{x}\!=\!(\vec{q},\vec{p})$.

The $\delta$-function in the density of encounters
(\ref{eq:density_of_encounters}) can be interpreted as classical
probability density for a trajectory starting at
$(\vec{q}_3,\vec{p}_3)$ to be at time $t'$ at a certain phase space
point which depends on $\vec{q}$, $\vec{p}$, $\vec{s}$, $\vec{u}$ and $t'$.
As the initial points $(\vec{q}_3,\vec{p}_3)$ are not located within
the encounter region, it is justified to utilize the ergodic property
of the chaotic system, which states that every accessible PS point is
equally likely to be reached by the classical dynamics.
We can thus approximate $\rho_{\alpha,\beta}$ by
\begin{equation}
  \rho_{\alpha,\beta}\to
  \frac{(2\pi\heff)^2}{\tenc\left(\vec{s},\vec{u}\right)}
  \frac{\delta^2
    \begin{pmatrix}
      \mathcal{H}^{\rmcl} (\vec{x}_3) -\mathcal{H}^{\rmcl}(\vec{x})\\
      \mathcal{N}(\vec{x}_3) -\mathcal{N}(\vec{x})
    \end{pmatrix}}
  {\Sigma(\vec{x})}\,,
\end{equation}
where $\Sigma(\vec{x})$ is the volume of the chaotic PS submanifold,
\begin{equation}
  \Sigma(\vec{x})=
  \int d^{2n} \!x'\,
  \delta^2
  \begin{pmatrix}
    \mathcal{H}^{\rmcl}(\vec{x}') - \mathcal{H}^{\rmcl}(\vec{x})\\
    \mathcal{N}(\vec{x}') - \mathcal{N}(\vec{x})
  \end{pmatrix}\,.
\end{equation}
Together with the integration over initial PS points $\vec{x}_3$,
ergodic PS averages are introduced, which lead to the following
substitution of initial momenta and final position:
\begin{equation}
  \left(p_{3,i}-p_i \right)^2
  \fin{q}_{j}\!\left(\vec{x}_3;t\right) \to
  \Braket{\left(p'_i-p_i\right)^2\fin{q}_j\!\left(
      \vec{x}';t\right)}_{\vec{x}}\,,
\end{equation}
where the ergodic PS average is defined as
\begin{equation}
  \Braket{f(\vec{x}')}_{\vec{x}}
  =\frac{
    \int d^{2n}\! x'\,
    \delta^2
    \begin{pmatrix}
      \mathcal{H}^{\rmcl}(\vec{x}') - \mathcal{H}^{\rmcl}(\vec{x})\\
      \mathcal{N}(\vec{x}') - \mathcal{N}(\vec{x})
    \end{pmatrix}
    f(\vec{x}')}{\Sigma(\vec{x})}\,.
  \label{eq:def_ergodic_PS_average}
\end{equation}
For times longer than the ergodic time $\lambda^{-1}$ we can further
assume that the final position is independent of its starting point,
and the average factorizes,
\begin{equation}
  \Braket{\left(p'_i-p_i\right)^2 \fin{q}_j\!\left(
  \vec{x}';t\right)}_{\vec{x}} =
  \Braket{\left(p'_i-p_i\right)^2}_{\vec{x}}\Braket{q'_j}_{\vec{x}}\,.
  \label{eq:ergodic_PS_average}
\end{equation}
With the same reasoning, we can also approximate the remaining factor
$\fin{q}_j\!\left( \vec{x};t\right)$ by its ergodic average
$\braket{q'_j}_{\vec{x}}$.
After introducing the Wigner function,
\begin{equation}
  W(\vec{q},\vec{p})
  \! =\!
  \int \! \frac{\rmd^n y}{(2\pi\heff)^n}
  \Psi^*\left(\vec{q}+\frac{\vec{y}}{2}\right)
  \Psi\left(\vec{q}-\frac{\vec{y}}{2}\right)
  \rme^{\frac{\rmi}{\heff}\vec{p}\vec{y}}\,,
\end{equation}
we see that the contribution of four-leg encounters can be written in
terms of a PS average weighted with the Wigner function,
\begin{equation}
  C^{(\mathrm{4le})}(t)
  =\int\! \rmd^n q\!
  \int\! \rmd^n p\,
  W\left(\vec{q},\vec{p}\right)
  I^{(\mathrm{4le})}\left(\vec{q},\vec{p};t\right)\,.
  \label{eq:4le_PS_average}
\end{equation}
Here the PS function
\begin{equation}
  I^{(\mathrm{4le})}\left(\vec{q},\vec{p};t\right)=
    \Braket{\left( p'_i-p_i \right)^2}_{\vec{x}}
    \Braket{q'_j}_{\vec{x}}^2 F^{(\mathrm{4le})}(t)
  \label{eq:4le_PS_function}
\end{equation}
contains the ergodic PS averages (\ref{eq:ergodic_PS_average}) and the
encounter integral
\begin{align}
  F^{(\mathrm{4le})}(t)
  &=\frac{1}{(2\pi\heff)^{n-2}}
  \int_{-c}^c\!\! \rmd^{n-2} s \!
  \int_{-c}^c\!\! \rmd^{n-2} u\,
  \rme^{ \frac{\rmi}{\heff}\vec{s}\vec{u}}
  \nonumber\\
  &\qquad\times
  \frac{t-\tenc(\vec{s},\vec{u})}{\tenc(\vec{s},\vec{u})}
    \Theta\left[t-\tenc(\vec{s},\vec{u})\right]\,.
    \label{eq:4le_encounter_integral}
\end{align}
In order to resolve the $\max$-function in the definition of $\tenc$,
we split the integrations over $\vec{s}$, $\vec{u}$.
This leads to the summation
\begin{equation}
  F^{(\mathrm{4le})}(t)\! =\! \sum_{i,j=1}^{n-2}F^{(\mathrm{4le})}_{ij}(t)\,,
\end{equation}
where
\begin{widetext}
  \begin{align}
    F^{(\mathrm{4le})}_{ij}(t)
    &=\frac{1}{(2\pi\heff)^{n-2}} \int_{-c}^c\!\!\rmd s_i\!
      \int_{-c}^c\!\!\rmd u_j\,
      \frac{t-\tenc(s_iu_j)}{\tenc(s_iu_j)}
      \Theta\left[t-\tenc(s_iu_j)\right]
      \left(\prod_{\substack{k,k'=1\\k\neq i,\,k'\neq j}}^{n-2}
    \!\!\int_{-|s_i|}^{|s_i|}\!\!\rmd s_k\!
    \int_{-|u_j|}^{|u_j|}\!\!\rmd u_{k'}\!\right)
    \rme^{ \frac{\rmi}{\heff}\vec{s}\vec{u}}\,,
  \end{align}
\end{widetext}
\noindent
with the encounter time
$\tenc(s_iu_j) \!=\! (1/\lambda)\log(c^2/|s_iu_j|)$, see
Eq.~(\ref{eq:encounter_time}).

One has to distinguish the cases $i\!\neq\! j$ from $i\!=\!j$ in order to
correctly interpret the product over $k$, $k'$.
The integrations over $s_k$ and $u_{k'}$ are easily performed, either
by a simple integration of an exponential for $k\!=\!j$, $k'\!=\!i$,
$i\!\neq\! j$, or by sorting the products such that $k\!=\!k'$ and using
Eq.~(\ref{eq:encint_0}).
For the integration over $s_i$, $u_j$, one first transforms the
integration over the subinterval $[-c,0]$ to $[0,c]$ by inverting the
sign of the integration variables. For the resulting integrations over
positive $s_i$, $u_j$ we then use the variable transformation
\cite{BroRahPRB2006}
\begin{align}
    (s_i,u_j) &\to (S,\sigma)
    = \left(\frac{s_iu_j}{c^2}, \frac{c}{u_j}\right)
    \textrm{, with } \left|\frac{\partial (s_i,u_j)}{\partial
        (S,\sigma)}\right| = \frac{c^2}{\sigma}
    \nonumber\\
    &\quad \textrm{ and }
    0<S<1,\, 1<\sigma<\frac{1}{S}\,.
\end{align}
The integration over $\sigma$ leads to the cancellation of
$\tenc(s_iu_j)$ in the denominator.
The argument of the Heaviside step function $\Theta$ demands
$t\!>\!(1/\lambda)\log(S^{-1}) $, which is equivalent to
$S \!>\! \exp(-\lambda t)$, thus raising the lower integration limit for
$S$.
We get as result for $i\!=\!j$
\begin{align}
    F^{(\mathrm{4le})}_{ii}(t)
    &= \left(\frac{2}{\pi}\right)^{n-2}
    \int_{\rme^{-\lambda t}}^1\rmd S
      \left(\lambda t -\log\left(\frac{1}{S}\right)\right)
      \nonumber\\
    &\qquad\times \Si^{n-3}\left(\frac{c^2S}{\heff}\right)
    \cos\left(\frac{c^2S}{\heff}\right)
      \frac{c^2}{\heff}\,,
\end{align}
and for $i\!\neq\! j$
\begin{align}
  F^{(\mathrm{4le})}_{ij}(t)
    &=\left(\frac{2}{\pi}\right)^{n-2}
    \int_{\rme^{-\lambda t}}^1\rmd S
      \left(\lambda t -\log\left(\frac{1}{S}\right)\right)
      \nonumber \\
    &\qquad\times \Si^{n-4}\left(\frac{c^2S}{\heff}\right)
      \frac{\sin^2\left(\frac{c^2S}{\heff}\right)}{S}\,,
\end{align}
where $\Si(z)=\int_0^z\rmd z' (\sin(z')/ z')$ denotes the sine
integral.

We can now perform the summation over indices $i$, $j$ to obtain an
integral expression for $F^{(\mathrm{4le})}$.
Note that
\begin{align}
    &(n-2)(n-3)\Si^{n-4}\left(\frac{c^2S}{\heff}\right)
      \frac{\sin^{2}\left(\frac{c^2S}{\heff}\right)}{S}
  \nonumber\\
    &\quad + (n-2) \Si^{n-3}\left(\frac{c^2S}{\heff}\right)
    \cos\left(\frac{c^2S}{\heff}\right)
      \frac{c^2}{\heff}
  \nonumber\\
   & = \frac{\rmd }{\rmd S } S \frac{\rmd}{\rmd S}
   \Si^{n-2}\left(\frac{c^2S}{\heff}\right)\,,
\end{align}
and thus
\begin{align}
    F^{(\mathrm{4le})}(t)
    &=\left(\frac{2}{\pi}\right)^{n-2}\int_{\rme^{-\lambda t}}^1\rmd S
      \left(\lambda t -\log\left(\frac{1}{S}\right)\right)
  \nonumber\\
    &\qquad\times
    \frac{\rmd }{\rmd S } S \frac{\rmd}{\rmd S}
    \Si^{n-2}\left(\frac{c^2S}{\heff}\right)\,.
\end{align}
The outer derivative in the second line is shifted to the first factor
in the integrand.
As $\rmd/\rmd S\,(t\!-\!\log(S^{-1})) \! = \! 1/S $ cancels the factor $S$, the
remaining integral is easily performed. To obtain more physical
insight at this stage, it is worth to introduce the Ehrenfest time,
\begin{equation}
  \tauE=\frac{1}{\lambda}\log\left(\frac{c^2}{\heff}\right)
  \quad\Leftrightarrow\quad
  \frac{c^2}{\heff}=\rme^{\lambda \tauE}\,.
  \label{eq:Ehrenfest_time}
\end{equation}
It is the time scale for which under hyperbolic dynamics
details of the order of $\heff$ can grow to the typical classical
action $c^2$. Using this, we obtain as final result
\begin{align}
  &F^{(\mathrm{4le})}(t)
    = \left(\frac{2}{\pi}\right)^{n-2}\hspace{-2ex}
    \lambda t (n-2)
    \Si^{n-3}\left(\rme^{\lambda \tauE}\right)
    \sin\left(\rme^{\lambda \tauE}\right)\nonumber\\
  &\quad -
    \left(\frac{2}{\pi}\right)^{n-2}\left[
    \Si^{n-2}\left(\rme^{\lambda \tauE}\right)
    -\Si^{n-2}\left(\rme^{\lambda\left(\tauE-t\right)}\right)\right]\,.
    \label{eq:4le_final_result}
\end{align}
In the semiclassical limit $\heff\!\ll\! c^2$, $\tauE$ in
Eq.~(\ref{eq:Ehrenfest_time}) is large compared to the ergodic time
$\lambda^{-1}$ implying a separation of time scales for the
OTOC. Thus, $\exp(\lambda\tauE)\!\gg\! 1$, and this has several
consequences:
\begin{itemize}
\item $\sin[\exp(\lambda \tauE)]$ is highly oscillatory and
  can be neglected in the phase space average
  (\ref{eq:4le_PS_average}).
\item $\Si[\exp(\lambda \tauE)]$ is well approximated by the
  asymptotic limit of the sine integral for large, positive arguments,
  $\Si[\exp(\lambda \tauE)]\!\approx\! \frac{\pi}{2}$.
\item For $t<\tauE$ Taylor-expansion around $t/\tauE\!=\!0$ yields
  \begin{align}
    &\Si^{n-2}\left(\rme^{\lambda \tauE}\right)
      -\Si^{n-2}\left(\rme^{\lambda (\tauE-t)}\right)\nonumber\\
    & \approx (n-2) \Si^{n-3} \left(\rme^{\lambda \tauE}\right)
      \sin\left(\rme^{\lambda \tauE}\right)\lambda t\,,
  \end{align}
  where the term linear in $t$ is the same highly oscillatory term as
  in the first item and can be neglected.
  (Alternatively, if not neglected, it would exactly cancel the
  oscillatory term for small $t$.)
\item For $t \!>\! \tauE$ we have
  $\exp[ \lambda (\tauE \!-\! t )] \!\ll\! 1$, and thus, by
  Taylor-expanding $\Si(y)$ around $y \!=\! 0$, we get
  \begin{align}
    \Si^{n-2}\left(\rme^{\lambda (\tauE-t)}\right)
    \approx \rme^{(n-2)\lambda (\tauE-t)}\,,
  \end{align}
  which is exponentially fast decaying for $t \!>\! \tauE$ and can be
  neglected for $t\!\gg\!\tauE$.
\end{itemize}
Combining the above considerations, we can well approximate
\begin{equation}
  F^{\mathrm{(4le)}}(t)
  \approx\left\{
    \begin{matrix}
      0    &\textrm{ if } t\ll \tauE
      \\
      -1
      &\textrm{ if } t\gg \tauE
    \end{matrix}\right\}
  \approx
  -\Theta\left(t-\tauE\right)\,.
\end{equation}
Hence the diagram class of the 4-leg-encounters only contributes after
a certain minimal time, the Ehrenfest time $\tauE$.
It is after this time that a description solely based on classical
dynamics breaks down, as interference contributions due to trajectory
constellations with encounter regions with an action difference of the
order $\heff$ start to exist.

\subsection{Contributions of 2-leg-encounters}
2-leg encounter diagrams are characterized by an encounter region that
contains either the starting or the end points of the quadruplet of
trajectories, see Fig.~\ref{fig:OTOC_diagram_classes} (b) and (c).
\subsubsection{Encounter at the beginning}
We start with diagram (b). Its contribution $C^{\textrm{(2le,(b))}}$
is calculated from a similar expression as $C^{\textrm{(4le)}}$,
Eq.~(\ref{eq:4le_starting_formula}), however with three major
differences:
\begin{itemize}
\item As the encounter region is at the beginning, the integration over
  $t'$ is over the interval $[0,\ts(\vec{s})]$.
\item The time of the encounter is reduced to
  $\tenc(t',\vec{u})\! =\! t'\! +\! \tu(\vec{u})$.
\item The difference of initial momenta is expressed through
  Eq.~(\ref{eq:initial_momentum_difference}) and has to be considered
  in the integration over $\vec{s}$.
\end{itemize}
Apart from a different treatment of the density $\rho_{\alpha,\beta}$,
which here can be directly used to cancel the integration over
$\vec{x}_3$, we apply the same steps which led to
Eq.~(\ref{eq:4le_PS_average}) for the 4-leg encounter.
Formally we arrive at the same PS average as in
Eq.~(\ref{eq:4le_PS_average}).
However, in this case the average is taken over the PS function
\begin{align}
  &I^{(\textrm{2le,(b)})}(\vec{q},\vec{p};t)\nonumber\\
  &=
    \Braket{q'_j}_{\vec{x}}^2
    \sum_{l,l'=1}^{n-2}
    \left[\veces{l}{\vec{x}}\right]_{p_i}
    \left[\veces{l'}{\vec{x}}\right]_{p_i}
    F_{ll'}^{(\mathrm{2le,(b)})}(t)
    \label{eq:2le_b_PS_average}
\end{align}
where the encounter integral reads
\begin{align}
  F_{ll'}^{(\mathrm{2le,(b)})}(t)
  &=\frac{1}{(2\pi\heff)^{n-2}}
    \int_{-c}^c\!\!\rmd^{n-2} s\! \int_{-c}^c \!\! \rmd^{n-2} u\,
    \rme^{\frac{\rmi}{\heff}\vec{s}\vec{u}}s_ls_{l'}
    \nonumber\\
  &\qquad\times
    \int_{0}^{\ts(\vec{s})}\hspace{-2ex}\rmd t'\,
    \frac{\Theta\left[t-\tenc(t',\vec{u})\right]}
    {\tenc(t',\vec{u})}\rme^{2\lambda t'}\,.
    \label{eq:2le_b_encounter_integral}
\end{align}
For $l\!\neq\! l'$ we immediately get $F_{ll'}^{(\mathrm{2le,(b)})}(t)\!=\!0$,
as the variable transformation $(s_l,u_l)\!\to\! -(s_l,u_l)$ results in
$F_{ll'}^{(\mathrm{2le,(b)})}(t)\! = \!-F_{ll'}^{(\mathrm{2le,(b)})}(t)$.
Thus only the case $l\!=\!l'$ needs to be considered.

We again split
\begin{equation}
  F_{ll}^{(\mathrm{2le,(b)})}(t)=\sum_{i,j=1}^{n-2}
  F_{l,ij}^{(\mathrm{2le,(b)})}(t)\,,
  \label{eq:2le_b_splitting}
\end{equation}
where
\begin{widetext}
  \begin{equation}
    F_{l,ij}^{(\mathrm{2le,(b)})}(t)
    = \frac{1}{(2\pi\heff)^{n-2}}
    \int_{-c}^c\!\!\rmd s_i\!
    \int_{-c}^c\!\!\rmd u_j\!
    \int_0^{\ts(s_i)}\hspace{-2ex}\rmd t'\,
    \frac{\Theta\left[t-\tenc(t',u_j)\right]}{\tenc(t',u_j)}
    \left(\prod_{\substack{k,k'=1\\k\neq i,\,k'\neq j}}^{n-2}\!\!
      \int_{-|s_i|}^{|s_i|}\!\!\rmd s_k\!
      \int_{-|u_j|}^{|u_j|}\!\!\rmd u_{k'}\right)
    s_l^2\rme^{ \frac{\rmi}{\heff}\vec{s}\vec{u}} \rme^{2\lambda t'},
  \end{equation}
\end{widetext}
\noindent
with $\ts(s_i)\!=\!(1/\lambda)\log(c/|s_i|)$ and the encounter time
$\tenc(t',u_j)\!=\!t'\! +\! (1/\lambda)\log(c/|u_j|)$.
For correctly resolving the products, we must again distinguish the
cases $i\!=\!j$ from $i\!\neq\! j$, and moreover the cases, when $l$
happens to be one of the indices $i$, $j$.
Using Eqs.~(\ref{eq:encint_0}, \ref{eq:encint_s2}), the integrations
over $s_k$, $u_{k'}$ for $k\!\neq\! i$, $k'\!\neq\! j$ are readily
performed.
For the last integrals we use the transformation
\cite{BroRahPRB2006}
\begin{align}
  (s_i,u_j,t')\to (T,S,\sigma)
  &= \left(  t' + \tu(u_j), \frac{s_iu_j}{c^2}, \frac{c}{u_j} \right)\,,
    \nonumber\\
  \textrm{ with }\left|
  \frac{\partial (s_i,u_j,t')}
  {\partial(T,S,\sigma)}\right|
  &=\frac{c^2}{\sigma}\nonumber\\
  \textrm{and } 0\leq T<\infty\,,\, 0
  & \leq S\leq\rme^{-\lambda T}\,,\,
    1\leq \sigma\leq\frac{1}{S}\,.
  \label{eq:2le_transformation}
\end{align}
The integration over $\sigma$ leads to a cancellation of the encounter
time in the denominator. The Heaviside step function transforms to
$\Theta(t\!-\!T)$, which introduces an upper bound in the integration over
$T$.

The results have the common structure
\begin{align}
  F_{l,ij}^{(\mathrm{2le,(b)})}(t)
  =\!\left(\frac{2}{\pi}\right)^{n-2}\hspace{-2ex}c^2\lambda
  \int_0^t\!\rmd T\, \rme^{2\lambda T}\!\!\int_0^{\rme^{-\lambda T}}
  \hspace{-4ex}\rmd S f_{l,ij}(S)\,,
  \label{eq:2le_common_structure}
\end{align}
and we must distinguish the following five cases:
\begin{itemize}
\item for $i\!=\!j\!=\!l$:
  \begin{equation}
    f_{l,ll}(S)
    = \Si^{n-3}\left(\frac{c^2S}{\heff}\right)
    \cos\left(\frac{c^2S}{\heff}\right)
    S^2\frac{c^2}{\heff}\,,
  \end{equation}
\item for $i\!=\!j\!\neq\! l$:
  \begin{align}
    f_{l,ii}(S)
    &= \Si^{n-4}\left(\frac{c^2S}{\heff}\right)
      \cos\left(\frac{c^2S}{\heff}\right)\nonumber\\
    &\quad \times \left(\frac{\heff}{c^2}\sin\left(\frac{c^2S}{\heff}\right)
      -S \cos\left(\frac{c^2S}{\heff}\right)
      \right)\,,
  \end{align}
\item for $i\! \neq\! j$, $i,j\!\neq\!l$:
  \begin{align}
    f_{l,ij}(S)
    &= \Si^{n-5}\left(\frac{c^2S}{\heff}\right)
      \frac{\sin^2\left(\frac{c^2S}{\heff}\right)}{S}\\
    &\quad \times \frac{\heff}{c^2}
      \left(\frac{\heff}{c^2}\sin\left(\frac{c^2S}{\heff}\right) - S
      \cos\left(\frac{c^2S}{\heff}\right) \right)\,,\nonumber
  \end{align}
\item for $i\! \neq\! j$, $i\!=\!l$:
  \begin{equation}
    f_{l,lj}(S)
    = \Si^{n-4}\left(\frac{c^2S}{\heff}\right)
    S \sin^2\left(\frac{c^2S}{\heff}\right)\,,
  \end{equation}
\item for $i\! \neq\! j$, $j\!=\!l$:
  \begin{align}
    f_{l,il}(S)
    &= \Si^{n-4}\left(\frac{c^2S}{\heff}\right)
      \frac{\sin\left(\frac{c^2S}{\heff}\right)}{S}
      \Bigg(
      2\frac{\heff}{c^2}S\cos\left(\frac{c^2S}{\heff}\right)
      \nonumber\\
    &\qquad\quad\left.+
      \left(S^2-2\left(\frac{\heff}{c^2}\right)^2\right)
      \sin\left(\frac{c^2S}{\heff}\right)
      \right)\,.
  \end{align}
\end{itemize}
The sum, Eq.~(\ref{eq:2le_b_splitting}), over all indices to obtain
$F_{ll}^{(\mathrm{2le,(b)})}(t)$ directly translates to a summation of
$f_{l,ij}(x)$ via Eq.~(\ref{eq:2le_common_structure}).
The latter sum can be conveniently rewritten as
\begin{align}
  \sum_{i,j=1}^{n-2} f_{l,ij}(S)
  = - \frac{\rmd}{\rmd S} S^3 \frac{\rmd}{\rmd S}
  \Si^{n-3}\left(\frac{c^2S}{\heff}\right)\Si''\left(\frac{c^2S}{\heff}\right).
\end{align}
This  identity allows one to easily perform the remaining
integrals over $S$ and $T$.
We obtain
\begin{align}
  F_{ll}^{(\mathrm{2le,(b)})}(t)
  & = -\left(\frac{2}{\pi}\right)^{n-2}\hspace{-2ex}c^2
    \Big[\Si^{n-3}\left(\rme^{\lambda\tauE}\right)
    \Si''\left(\rme^{\lambda\tauE}\right)\nonumber
  \\
  &\quad -
    \Si^{n-3}\left(\rme^{\lambda(\tauE-t)}\right)
    \Si''\left(\rme^{\lambda(\tauE-t)}\right)\Big]\,.
    \label{eq:2le_b_result}
\end{align}
The result contains the second derivative of $\Si$,
$\Si''(z)\!=\!\cos(z)/z\! -\! \sin(z)/z^2$, which contains oscillatory
functions. We consider again the limiting cases:
\begin{itemize}
\item For $t\!\ll\!\tauE$ we get
  $\exp[\lambda(\tauE\!-\!t)]\! \approx \!\exp(\lambda\tauE)$, and
  $\Si''[ \exp(\lambda\tauE)]$ only contains highly oscillatory
  factors, which we can neglect in the semiclassical limit
  $\heff\!\ll\! c^2$.
\item For $t\!\gg\!\tauE$ we expand around
  $\exp[\lambda(\tauE\!-\!t)]\!\approx\! 0$
  \begin{align}
  &\Si^{n-3}\left(\rme^{\lambda(\tauE-t)}\right)
    \Si''\left(\rme^{\lambda(\tauE-t)}\right)\nonumber\\
  &\approx
    \Si'''(0) \rme^{(n-2)\lambda(\tauE-t)},
  \end{align}
  where $\Si'''(0)\!=\!-5/3$. As for the 4-leg encounter, this
  contribution is exponentially small.
\end{itemize}
For times $t\!\ll\!\tauE$ and $t\!\gg\!\tauE$ the diagrams in
Fig.~\ref{fig:OTOC_diagram_classes} (b) are negligible in the
semiclassical limit.
Only for $t\!\approx\!\tauE$, the above terms can, in
principle, produce non-negligible contributions.
However, for these times the results depend on the (sharp) cutoff
value $c$ of the encounter integrations, indicating that the
quantitative result of the encounter integration is not very
meaningful.
However, qualitatively, our results indicate that the interference
mechanism behind diagram (b) accounts, together with other diagrams,
for the smooth crossover between the pre- and post-Ehrenfest time
behavior of OTOCs.

\subsubsection{Encounter at the end}
We now turn to the related 2-leg encounter class of diagram (c) in
Fig.~\ref{fig:OTOC_diagram_classes}, where the final points of the
quadruplet of trajectories is contained inside the encounter.
In this case, the following modifications to
Eq.~(\ref{eq:4le_starting_formula}) are required:
\begin{itemize}
\item The integration interval for $t'$ is
  $[t\!-\!\tu(\vec{u}),t]$.
\item The encounter time is
  $\tenc(t',\vec{s})=\ts(\vec{s})\!+\!(t\!-\!t')$.
\item The product of final positions is expressed through
  Eq.~(\ref{eq:final_position_difference}).
  This leads to correlated final points $\braket{{q'_j}^2}_{\vec{x}}$
  in the ergodic average, and to a corresponding modification in the
  integration over $\vec{u}$.
\end{itemize}
The contributions are calculated by
\begin{widetext}
  \begin{equation}
    I^{(\textrm{2le,(c)})}(\vec{q},\vec{p};t)
    =
    \Braket{(p'_i-p_i)^2}_{\vec{x}}\left(
      \Braket{{q'_j}^2}_{\vec{x}}
      F^{(\mathrm{2le,(c)})}(t)
      -\frac{1}{2}
      \sum_{l,l'=1}^{n-2}\!
      \left[\veceu{l}{\vecfin{x}(\vec{x};t)}\right]_{q_j}\!\!
      \left[\veceu{l'}{\vecfin{x}(\vec{x};t)}\right]_{q_j}\!\!
      F_{ll'}^{(\mathrm{2le,(c)})}(t)
    \right)\,,
    \label{eq:2le_c_PS_average}
  \end{equation}
  where
  \begin{align}
    F^{(\mathrm{2le,(c)})}(t)
    &=\frac{1}{(2\pi\heff)^{n-2}}
      \int_{-c}^c\!\!\rmd^{n-2} s\!
      \int_{-c}^c  \!\! \rmd^{n-2} u\,
      \rme^{ \frac{\rmi}{\heff}\vec{s}\vec{u}}
      \int_{t-\tu(\vec{u})}^{t}\hspace{-3ex}\rmd t'\,
      \frac{\Theta\left[t-\tenc(t',\vec{s})\right]}
      {\tenc(t',\vec{s})}\,,
      \label{eq:2le_c_1_encounter_integral}
    \\[2ex]
    F_{ll'}^{(\mathrm{2le,(c)})}(t)
    &=\frac{1}{(2\pi\heff)^{n-2}}\!
      \int_{-c}^c\!\!\! \rmd^{n-2} s\! \int_{-c}^c\! \!\! \rmd^{n-2} u\,
      \rme^{ \frac{\rmi}{\heff}\vec{s}\vec{u}}u_lu_{l'}
      \int_{t-\tu(\vec{u})}^{t}\hspace{-3ex}\rmd t'\,
      \frac{\Theta\left[t-\tenc(t',\vec{s})\right]}
      {\tenc(t',\vec{s})}\rme^{2\lambda (t-t')}\,.
      \label{eq:2le_c_2_encounter_integral}
  \end{align}
\end{widetext}
In a first step, we interchange the variable names for stable and
unstable coordinates, $\vec{s}\!\leftrightarrow\!\vec{u}$, which formally
interchanges $\ts(\vec{s})\!\leftrightarrow\! \tu(\vec{u})$.
Then we perform a variable transformation $t'\!\rightarrow\! t\!-\!t'$, which
inverts the arrow of time.
These steps transform the calculations for an encounter at the end to
those for an encounter at the beginning of the trajectories, and we
immediately obtain
$F_{ll'}^{(\mathrm{2le,(c)})}(t) \!=\!
F_{ll'}^{(\mathrm{2le,(b)})}(t)$.
It thus remains to calculate $F^{(\mathrm{2le,(c)})}(t)$, which in the
transformed version reads
\begin{align}
  F^{(\mathrm{2le,(c)})}(t)
  &=\frac{1}{(2\pi\heff)^{n-2}}
    \int_{-c}^c\!\! \rmd^{n-2} s
    \int_{-c}^c\!\! \rmd^{n-2} u\,
    \rme^{ \frac{\rmi}{\heff}\vec{s}\vec{u}}
    \nonumber\\
  &\qquad\times\int_{0}^{\ts(\vec{s})}\hspace{-2ex}\rmd t'\,
    \frac{\Theta\left[t-\tenc(t',\vec{u})\right]}
    {\tenc(t',\vec{u})}.
\end{align}
In the same spirit as for 4-leg encounter diagrams in the previous
section, we write
\begin{equation}
  F^{(\mathrm{2le,(c)})}(t)=\sum_{i,j=1}^{n-2}F_{ij}^{(\mathrm{2le,(c)})}(t)
\end{equation}
to resolve the $\max$-functions inherent in $\ts(\vec{s})$ and
$\tu(\vec{u})$ in Eqs.~(\ref{eq:stable_unstable_times}), and
finally use Eq.~(\ref{eq:2le_transformation}) to
transform the last integrals.
We obtain
\begin{equation}
  F_{ij}^{(\mathrm{2le,(c)})}(t)
  =\frac{1}{(2\pi\heff)^{n-2}}\lambda
  \int_0^t\!\rmd T\! \int_0^{\rme^{-\lambda T}}\hspace{-3ex} \rmd S\,
  f_{ij}^{(\mathrm{2le,(c)})}(S)\,,
\end{equation}
where
\begin{itemize}
\item for $i\!=\!j$:
  \begin{equation}
    f_{ii}^{(\mathrm{2le,(c)})}(S)
    =  \Si^{n-3}\left(\frac{c^2S}{\heff}\right)
    \cos\left(\frac{c^2S}{\heff}\right)
    \frac{c^2}{\heff}\,,
  \end{equation}
\item for $i\!\neq\! j$:
  \begin{equation}
    f_{ij}^{(\mathrm{2le,(c)})}(t)
    =  \Si^{n-4}\left(\frac{c^2S}{\heff}\right)
    \frac{\sin^2\left(\frac{c^2S}{\heff}\right)}{S}\,.
  \end{equation}
\end{itemize}
After summation over indices, we find
\begin{equation}
  \sum_{i,j=1}^{n-2} f_{ij}^{(\mathrm{2le,(c)})}(S) =
  \frac{\rmd}{\rmd S}S\frac{\rmd}{\rmd S}
  \Si^{n-2}\left(\frac{c^2S}{\heff}\right),
\end{equation}
which allows us to easily evaluate the final integrals.
We eventually obtain
\begin{equation}
  F^{(\mathrm{2le,(c)})}(t) =\! \left(\frac{2}{\pi}\right)^{n-2}\!\!
  \left[\Si^{n-2}\left(\rme^{\lambda \tauE }\right)
    - \Si^{n-2}\left(\rme^{\lambda(\tauE-t }\right)\!\right]\,.
  \label{eq:2le_c_final_result}
\end{equation}
Following the same arguments as in the section about the
4-leg-encounter, this term is only contributing for times larger than
the Ehrenfest time $\tauE$ and can be approximated by
$F^{(\mathrm{2le,(c)})}(t) \!\approx\! \Theta(t\!-\!\tauE)$ in the
semiclassical limit.
Both, the diagram (c) and (d) in Fig.~\ref{fig:OTOC_diagram_classes}
contribute to the OTOC for $t\!>\!\tauE$.

\subsection{Contributions of 0-leg-encounters}
In this section we calculate the contribution $C^{(\mathrm{0le})}(t)$
shown in Fig.~\ref{fig:OTOC_diagram_classes} (a), where the quadruplet
of trajectories is fully contained within an encounter,
\emph{i.e.}~the trajectories stay close to each other for the whole
time.
The starting point for the calculation of $C^{(\mathrm{0le})}(t)$
differs from the one of $C^{(\mathrm{4le})}$,
Eq.~(\ref{eq:4le_starting_formula}), in the following items (see also
\cite{Waltner2012,BroRahPRB2006b})
\begin{itemize}
\item As the encounter stretches over the full time $t$, the
  integration interval for $t'$ is $[0,t]$ and the encounter time is
  $\tenc\!=\!t$.
  There is no Heaviside step function $\Theta$ in time involved any
  more.
\item As the encounter time is fixed, the integration interval for the
  components of $\vec{s}$ is reduced to
  $[-c\exp(-\lambda t'),c\exp(-\lambda t')]$ to ensure none of the
  stable components grows larger than the maximal value $c$
  in the available time $t'$.
  With the same reasoning, the integration
  intervals for the components of $\vec{u}$ become
  $[-c\exp[-\lambda (t\!-\!t')],c\exp[-\lambda(t\!-\!t')]]$.
\item Both, the initial momenta difference and the product of final
  positions, have to be interpreted in view of
  Eqs.~(\ref{eq:initial_momentum_difference},
  \ref{eq:final_position_difference}) and be respected in the
  integrations over $\vec{s}$, $\vec{u}$ and when using ergodicity
  arguments.
\item The $\delta$-function in the density $\rho_{\alpha,\beta}$ of
  partner trajectories can again be directly used for canceling
  the integration over $\vec{x}_3$.
\end{itemize}
After the initial transformations, which convert the contribution into
a PS average, we arrive at
\begin{widetext}
  \begin{equation}
    I^{(\mathrm{0le})}(\vec{q},\vec{p};t)
    =\!\! \sum_{l,l'=1}^{n-2}\!\left[\veces{l}{\vec{x}}\right]_{p_i}\!\!
    \left[\veces{l'}{\vec{x}}\right]_{p_i}\!\!
    \left[\! \Braket{{q'_j}^2}_{\vec{x}}\!
      F_{ll'}^{(\mathrm{0le},1)}(t)
      -\frac{1}{2}\!\!\!
      \sum_{m,m'=1}^{n-2}\!\!
      \left[\veceu{m}{\vecfin{x}(\vec{x};t)}\right]_{q_j}\!\!
      \left[\veceu{m'}{\vecfin{x}(\vec{x};t)}\right]_{q_j}\!\!
      F_{ll'mm'}^{(\mathrm{0le},2)}(t)\right]\,,
    \label{eq:0le_starting_formula}
  \end{equation}
  with encounter integrals
  \begin{align}
    F_{ll'}^{(\mathrm{0le},1)}(t)
    & =\frac{1}{(2\pi\heff)^{n-2}}
      \int_0^t\rmd t'\,\rme^{2\lambda t'}
      \int_{-c\rme^{-\lambda t'}}^{c\rme^{-\lambda t'}}
      \hspace{-3ex} \rmd^{n-2}s\!
      \int_{-c\rme^{-\lambda (t-t')}}^{c\rme^{-\lambda (t- t')}}
      \hspace{-4ex} \rmd^{n-2}u \,
      \frac{\rme^{\frac{\rmi}{\heff}\vec{s}\vec{u}}}{t}
      s_ls_{l'}\,,
      \label{eq:0le_1_encounter_integral}
    \\[4ex]
    F_{ll'mm'}^{(\mathrm{0le},2)}(t)
    &=\frac{1}{(2\pi\heff)^{n-2}}
      \int_0^t\rmd t'\,\rme^{2\lambda t}
      \int_{-c\rme^{-\lambda t'}}^{c\rme^{-\lambda t'}}
      \hspace{-3ex} \rmd^{n-2}s\!
      \int_{-c\rme^{-\lambda (t-t')}}^{c\rme^{-\lambda (t- t')}}
      \hspace{-4ex} \rmd^{n-2}u \,
      \frac{\rme^{\frac{\rmi}{\heff}\vec{s}\vec{u}}}{t}
      s_ls_{l'}u_{m}u_{m'}\,.
      \label{eq:0le_2_encounter_integral}
  \end{align}
\end{widetext}
With the same reasoning as for 2-leg encounters, the integral
$F_{ll'}^{(\mathrm{0le},1)}(t)$ does not vanish for $l\!=\!l'$.
Using Eqs.~(\ref{eq:encint_0}, \ref{eq:encint_s2}) we calculate
\begin{equation}
  F_{ll}^{(\mathrm{0le},1)}(t)
  =-\left(\frac{2}{\pi}\right)^{n-2}\hspace{-3ex}c^2
  \Si^{n-3}\left(\rme^{\lambda(\tauE- t)}\right)
  \Si''\left(\rme^{\lambda(\tauE- t)}\right)\,.
\end{equation}
As has been argued after Eq.~(\ref{eq:2le_b_result}), this term
neither contributes in the case $t\!\ll\!\tauE$ nor for $t\!\gg\!\tauE$, but
qualitatively, the underlying interference mechanism is involved in
the crossover regime at $t\!\approx\!\tauE$.

For $F_{ll'mm'}^{(\mathrm{0le},2)}$ four indices are involved, and we find
three classes of non-vanishing integrals, which are treated using all
the Eqs.~(\ref{eq:encint_0}-\ref{eq:encint_su2}).
\begin{widetext}
  \begin{itemize}
  \item[(a)] For $l\!=\!l'$, $m\!=\!m'$, $l\!\neq\! m$ we get
    \begin{equation}
      F_{llmm}^{(\mathrm{0le},2)}(t)
      = \left(\frac{2}{\pi}\right)^{n-2}\hspace{-3ex}c^4
      \Si^{n-4}\left(\rme^{\lambda(\tauE-t)}\right)
      \left[\Si''\left(\rme^{\lambda(\tauE-t)}\right)\right]^2\!.
    \end{equation}
  \item[(b)] If the set of indices $\{l,l'\}\!=\!\{m,m'\}$ are equal
    without being all the same, \emph{i.e.}~$l\!\neq\! l'$, we get
    \begin{equation}
      F_{ll' mm'}^{(\mathrm{0le},2)}(t)
      = - \left(\frac{2}{\pi}\right)^{n-2}\hspace{-3ex}c^4
      \Si^{n-4}\!\left(\rme^{\lambda(\tauE- t)}\right)\rme^{2\lambda(t-\tauE)}
      \left[\Si\left(\rme^{\lambda(\tauE- t)}\right)-
        \sin\left(\rme^{\lambda(\tauE- t)}\right)
      \right]^2.
      \label{eq:0le_equal_sets}
    \end{equation}
  \item[(c)] If all indices are the same, we get
    \begin{align}
      F_{llll}^{(\mathrm{0le},2)}(t)
      &= -\left(\frac{2}{\pi}\right)^{n-2}\hspace{-3ex}c^4
        \Si^{n-4}\left(\rme^{\lambda(\tauE- t)}\right)\rme^{2\lambda(t-\tauE)}
        \Bigg[2
        \left[
        \Si\left(\rme^{\lambda(\tauE- t)}\right)
        -\sin\left(\rme^{\lambda(\tauE- t)}\right)
        \right]^2\nonumber\\[1ex]
      &\qquad   +\sin\left(\rme^{\lambda(\tauE- t)}\right)
        \Si\left(\rme^{\lambda(\tauE- t)}\right)
        -2
        \sin^2\left(\rme^{\lambda(\tauE- t)}\right)
        +\cos\left(\rme^{\lambda(\tauE- t)}\right)
        \Si\left(\rme^{\lambda(\tauE- t)}\right)\rme^{\lambda(\tauE- t)}\Bigg]
        \label{eq:0le_equal_indices}
    \end{align}
  \end{itemize}
\end{widetext}
Case (a) is multiplied with $\Si''[\exp[\lambda(\tauE\!-\! t)]]$ and thus,
like $F_{ll}^{(\mathrm{0le},1)}(t)$, can be neglected for $t\!\ll\!\tauE$
and $t\!\gg\!\tauE$.
For case (b) we have for $t\!\ll\!\tauE$:
\begin{equation}
  \Si\left(\rme^{\lambda(\tauE- t)}\right)-
  \sin\left(\rme^{\lambda(\tauE- t)}\right)
  \approx
  \Si\left(\rme^{\lambda \tauE}\right)
  \approx \frac{\pi}{2}\,,
\end{equation}
\emph{i.e.}~the highly oscillatory term
$\sin[\exp[\lambda(\tauE\!-\! t)]$ is neglected and we use
the asymptotic value for $\Si$.
For $t\!\gg\!\tauE$, we obtain from a Taylor expansion around
$\exp[\lambda(\tauE\!-\! t)]\!\approx\! 0$
\begin{equation}
  \Si\left(\rme^{\lambda(\tauE- t)}\right)-
  \sin\left(\rme^{\lambda(\tauE- t)}\right)
  \approx \frac{1}{9} \rme^{3\lambda(\tauE-t)}\,,
\end{equation}
and thus, as
$\Si[\exp[\lambda(\tauE\!-\! t)]]\!\approx\! \exp[\lambda(\tauE\!-\!t)]$,
\begin{equation}
  F_{ll' mm'}^{(\mathrm{0le},2)}(t)\approx
  -\frac{1}{81}
  \left(\frac{2}{\pi}\right)^{n-2}\hspace{-3ex}c^4\rme^{n\lambda(\tauE-t)}\,,
  \label{eq:exp_suppression_0le_b}
\end{equation}
\emph{i.e.}~the contribution becomes exponentially suppressed after
the Ehrenfest time.
We can thus approximate
\begin{align}
  F_{ll' mm'}^{(\mathrm{0le},2)}(t)
  &\approx
    - c^4\rme^{2\lambda(t-\tauE)}\Theta(\tauE-t)\nonumber\\
  & = -\heff^2\rme^{2\lambda t}\Theta(\tauE-t)\,.
\end{align}
Note that since we have $\{l,l'\} \!=\! \{m,m'\}$ we get an additional
combinatorial factor 2 when reducing the fourfold sum over
$l,l',m,m'$ in Eq.~(\ref{eq:0le_starting_formula}) to a twofold one
over $l,l'$ with $l\!\neq\! l'$.
The case of equal indices $l\!=\!l'$ is still excluded from this
summation, but using case (c), which also contains the same
contribution as case (b) (including the prefactor 2), we can complete
the summation.
It remains to discuss the additional terms in the last line of
Eq.~(\ref{eq:0le_equal_indices}).
Those can be neglected for $t\!\ll\!\tauE$ as they all contain highly
oscillatory factors. For $t\!\gg\!\tauE$ we find
\begin{align}
  &\sin\left(\rme^{\lambda(\tauE- t)}\right)
  \Si\left(\rme^{\lambda(\tauE- t)}\right)
  -2\sin^2\left(\rme^{\lambda(\tauE- t)}\right)\nonumber\\
  &\quad+\cos\left(\rme^{\lambda(\tauE- t)}\right)
    \Si\left(\rme^{\lambda(\tauE- t)}\right)\rme^{\lambda(\tauE- t)}\nonumber\\
  &\approx-\frac{1}{9}\rme^{4\lambda(\tauE- t)}\,.
\end{align}
This leads to a suppression of $F_{llll}^{(\mathrm{0le},2)}(t)$ for
$t\!\gg\!\tauE$, which is less strong than the one for
$F_{ll' mm'}^{(\mathrm{0le},2)}(t)$ in
Eq.~(\ref{eq:exp_suppression_0le_b}), but still exponential.
Thus, the overall exponential suppression reads
\begin{equation}
  F_{llll}^{(\mathrm{0le},2)}(t)
  \approx
  \frac{1}{9}\rme^{(n-2)\lambda(\tauE- t)}\,.
\end{equation}

\subsection{Summary}
In the previous subsections we found that diagram (a) in
Fig.~\ref{fig:OTOC_diagram_classes} fully describes, via
Eq.~(\ref{eq:0le_starting_formula}) together with
Eqs.~(\ref{eq:0le_equal_sets}, \ref{eq:0le_equal_indices}), the
dynamics of the OTOCs for $t\!<\!\tauE$.
The PS function $I_<^{(\mathrm{0le})}(\vec{q},\vec{p};t)$
corresponding to these results and used in the PS average,
Eq.~(\ref{eq:PS_average}), is found to be
\begin{equation}
  I_<(\vec{q},\vec{p};t)
  \approx\!\!
  \left(
    \sum_{l=1}^{n-2}\!
    \left[\veces{l}{\vec{x}}\right]_{p_i}\!\!
    \left[\veceu{l}{\vecfin{x}(\vec{x};t)}\right]_{q_j}\!\!\right)^2
  F_<(t)\,
  \label{eq:I_smaller}
\end{equation}
where the early-time exponential growth of OTOCs is contained in the
function
\begin{align}
  F_<(t)
  &= \left(\frac{2}{\pi}\right)^{n-2}\hspace{-3ex}c^4
    \rme^{2\lambda(t-\tauE)}
    \Si^{n-4}\!\left(\rme^{\lambda(\tauE- t)}\right)
    \nonumber\\
  &\qquad\times
    \left[\Si\left(\rme^{\lambda(\tauE- t)}\right)-
    \sin\left(\rme^{\lambda(\tauE- t)}\right)
    \right]^2\nonumber\\[1ex]
  &\approx \heff^2\rme^{2\lambda t}\Theta(\tau-t)\,.
\end{align}
For $t\!>\!\tauE$, the sum of contributions from diagrams (c) and (d)
produces the long-time saturation of OTOCs.
As seen from Eqs.~(\ref{eq:4le_PS_function},
\ref{eq:2le_c_PS_average}), with their temporal behavior given in
Eqs.~(\ref{eq:4le_final_result}, \ref{eq:2le_c_final_result}), their
combined contribution reads
\begin{equation}
  I_>\left(\vec{q},\vec{p};t\right) =
    \Braket{\left( p'_i-p_i \right)^2}_{\vec{x}}
  \left(\Braket{{q'_j}^2}_{\vec{x}}-\Braket{q'_j}_{\vec{x}}^2 \right)
  F_>(t)\,,
  \label{eq:I_larger}
\end{equation}
where
\begin{align}
  F_>(t)
  &=
    \left(\frac{2}{\pi}\right)^{n-2}
    \left[
    \Si^{n-2}\left(\rme^{\lambda \tauE}\right)
    -\Si^{n-2}\left(\rme^{\lambda\left(\tauE-t\right)}\right)\right]
    \nonumber\\[1ex]
     &\approx \Theta(t-\tauE)\,.
\end{align}

\section{Generalization to other operators}
The key ingredient for our method to understand OTOCs is to use
semiclassical techniques, which translate the quantum operators
$\hat{p}_i$ and $\hat{q}_j$ to their corresponding classical partners
while keeping the quantum mechanical phase information.
In the classical PS, we used the local linearization of Hamilton's
equations of motion to connect these classical functions to the
hyperbolic property of the chaotic system.
Furthermore, the ergodic property produced variances of these PS
functions.

In view of these points, a generalization of OTOCs to other operators,
$\braket{\Psi||[\hat{A},\hat{B}(t)]|^2|\Psi}$, appears to be
straightforward, if the following assumptions are fulfilled:
\begin{itemize}
\item The operators $\hat{A}$, $\hat{B}$ are smooth functions of the
  operators $\hat{q}_i$, $\hat{p}_i$, $i\!=\!1,\ldots,n$, in the sense
  that we can write $\hat{A}$, $\hat{B}$ as a sum of products of powers
  of position and momentum quadrature operators.
\item To avoid additional contributions to the overall action
  difference in the phase factor in
  Eq.~(\ref{eq:OTOC_sc_integral_representation}), the operators
  $\hat{A}$, $\hat{B}$ are not allowed to depend on $\heff^{-1}$.
  Hence, for instance, displacement operators
  $\exp(-(\rmi/\heff)y \hat{p}_i)$ would require a refined treatment.
\end{itemize}
With these assumptions, we expect our methods to apply.
The classical functions corresponding to the quantum operators are
constructed by replacing operators $\hat{q}_i$, $\hat{p}_i$ in the
expansion by the corresponding trajectory-based equivalents,
\emph{i.e.}~initial position and momentum quadratures in $\hat{A}$,
and final ones in $\hat{B}$.
Any dependence on powers of $\heff$ of single terms in these expansion
must be dropped as we are working in the leading order semiclassical
limit $\heff\!\ll\! c^2$.
These terms are expected to arise from different ordering of the
quantum operators $\hat{q}_i$, $\hat{p}_i$ and can be avoided from the
beginning by using operators and classical functions which are linked
to each other by the classical-quantum correspondence principle of
Weyl-symbols and Wigner transformations \cite{Schleich2001}.

Denoting the classical functions by $A(\vec{q},\vec{p})\!=\!A(\vec{x})$
and $B(\vec{x})$ it is straightforward to see that in the integrand of
Eq.~(\ref{eq:OTOC_sc_integral_representation}) we substitute
\begin{equation}
  \left(\init{p}_{\alpha',i}\!-\!\init{p}_{\alpha,i} \right)\fin{q}_{\alpha,j}
  \left(\init{p}_{\beta,i}\!-\!\init{p}_{\beta',i} \right)\fin{q}_{\beta,j}
\end{equation}
by
\begin{align}
  &\left[
    A\left(\vecinit{x}_{\alpha'}\right)
    \!-\!
    A\left(\vecinit{x}_{\alpha}\right)
    \right]\!
    B\left(\vecfin{x}_{\alpha}\right)
    \nonumber\\
  &\quad\times
    \left[
    A\left(\vecinit{x}_{\beta}\right)
    \!-\!
    A\left(\vecinit{x}_{\beta'} \right)
    \right]\!
    B\left(\vecfin{x}_{\beta}\right)
    \,.
\end{align}
For diagrams (b), (c) and (d) of Fig.~\ref{fig:OTOC_diagram_classes}
the beginning and/or the ends of the trajectories are not contained
inside an encounter region, and we approximate parts of the above
expression by their ergodic averages.
Note that as $\alpha'$, $\beta$ start at PS points which are
associated with the Wigner function in Eq.~(\ref{eq:PS_average}),
$A(\vecinit{x}_{\alpha'})$, $A(\vecinit{x}_{\beta})$ turn into
$A(\vec{x})$.
Like $p_i$ in Eq.~(\ref{eq:larger}) they are treated as constants in the
ergodic average Eq.~(\ref{eq:ergodic_PS_average}), but are later averaged
in the PS average, Eq.~(\ref{eq:PS_average}), involving the Wigner function.

For diagrams (a), (b) and (c) in Fig.~\ref{fig:OTOC_diagram_classes},
the initial and/or final points of the trajectories are contained
within an encounter, and thus we have to express the corresponding
functions through the local hyperbolic variables.
Equations (\ref{eq:initial_momentum_difference},
\ref{eq:final_position_difference}) are thus modified to
\begin{align}
  &A\left(\vecinit{x}_{\alpha'}\right)\!-\!A\left(\vecinit{x}_{\alpha}\right)
    \approx - \left[\frac{\partial A}{\partial
    \vec{x}}\left(\vecinit{x}_\beta\right)\right]
    \cdot\sum_{l=1}^{n} s_l\rme^{\lambda t'} \veces[\beta]{l}{0}\nonumber\\
  &A\left(\vecinit{x}_{\beta}\right)\!-\!A\left(\vecinit{x}_{\beta'}\right)
    \approx - \left[\frac{\partial A}{\partial
    \vec{x}}\left(\vecinit{x}_\beta\right)\right] \cdot\sum_{l=1}^{n}
    s_l\rme^{\lambda t'} \veces[\beta]{l}{0}\nonumber
  \end{align}
  \begin{align}
  &B\left(\vecfin{x}_{\alpha}\right)B\left(\vecfin{x}_{\beta}\right)\nonumber\\
  &\approx B^2\left(\vecfin{x}_{\beta}\right)
    -\frac{1}{2}
    \left(
    \left[
    \frac{\partial B}{\partial \vec{x}}\left(\vecfin{x}_\beta\right)
    \right]
    \cdot
    \sum_{l=1}^{n} u_l\rme^{\lambda(t- t')} \veceu[\beta]{l}{t} \right)^2
\end{align}
In view of our methods used, we note that this changes the PS function
$I(\vec{q},\vec{p};t)$, Eqs.~(\ref{eq:4le_PS_function},
\ref{eq:2le_b_PS_average}, \ref{eq:2le_c_PS_average},
\ref{eq:0le_starting_formula}), by using different ergodic averages
and adjusting the terms involving the stable and unstable
directions.
However, the encounter integrals $F(t)$ in
Eqs.~(\ref{eq:4le_encounter_integral},
\ref{eq:2le_b_encounter_integral},
\ref{eq:2le_c_1_encounter_integral},
\ref{eq:2le_c_2_encounter_integral},
\ref{eq:0le_1_encounter_integral}, \ref{eq:0le_2_encounter_integral})
remain the same.
The OTOC's result for operators fulfilling the above assumptions is
thus obtained by adjusting the classical information in the PS
functions $I_<$ and $I_>$ in Eqs.~(\ref{eq:I_smaller},
\ref{eq:I_larger}).

\section{Frequently used integrals in the calculations of encounters}
The following integrals are frequently obtained during the
calculations of encounter diagrams. Let $a,b$ be positive and
dimensionless real parameters.
$\Si(z) \!=\! \int_0^z\! \rmd z' (\sin(z')/z')$ defines the sine integral.
Then
\begin{align}
  \bullet
  &\int_{-a}^{a}\rmd s\int_{-b}^{b} \rmd u\, \rme^{\frac{\rmi}{\heff}su}
    = 4 \heff \Si\left(\frac{ab}{\heff}\right)\,,
    \label{eq:encint_0}
  \\[2ex]
  \bullet
  &\int_{-a}^{a}\rmd s\int_{-b}^{b} \rmd u\, s^2\rme^{\frac{\rmi}{\heff}su}
    = -4 \heff a^2 \Si''\left(\frac{ab}{\heff}\right)\nonumber\\
  &=4 \heff^3 \frac{1}{b^2} \sin\left(\frac{ab}{\heff}\right)
    -4 \heff^2 \frac{a}{b} \cos\left(\frac{ab}{\heff}\right),
    \label{eq:encint_s2}\\[2ex]
  \bullet
  &\int_{-a}^{a}\rmd s\int_{-b}^{b} \rmd u\, su \rme^{\frac{\rmi}{\heff}su}
    = - 4 \heff^2\rmi \left.\left(y^2
    \frac{\rmd}{\rmd y}
    \frac{\Si(y)}{y}\right)\right|_{y=\frac{ab}{\heff}}\nonumber\\
  & = \rmi 4 \heff^2\!
    \left[
    \Si\left( \frac{ab}{\heff} \right)
     - \sin\left( \frac{ab}{\heff} \right)
    \right],
    \label{eq:encint_su}\\[2ex]
  \bullet
  &\int_{-a}^{a}\rmd s\int_{-b}^{b} \rmd u\, s^2u^2 \rme^{\frac{\rmi}{\heff}su}
   \!=\! - 4 \heff^3\!
    \left.
    \left( y^3 \frac{\rmd^2}{\rmd y^2} \frac{\Si(y)}{y}\right)
    \right|_{y=\frac{ab}{h}}\nonumber\\[1ex]
  & = - 4 \heff^3\!
    \left[\!
    \frac{ab}{\heff}
    \cos\left( \frac{ab}{\heff} \right)\!
    \!-\! 3 \sin\left( \frac{ab}{\heff} \right)\!
    \!+\! 2 \Si\left( \frac{ab}{\heff} \right)\!\!
    \right].
    \label{eq:encint_su2}
\end{align}

%
\vspace*{\fill}

\end{document}